\documentclass[preprint2]{aa}
\usepackage{lscape}
\usepackage{graphicx}
\usepackage{float}
 \usepackage{caption}

\DeclareCaptionFormat{cont}{#1 (cont.)#2#3\par}

\usepackage{titlesec}

\begin{document}


\title {The observed chemical structure of L1544\thanks{Based on observations carried out with the IRAM 30m Telescope. IRAM is supported by INSU/CNRS (France), MPG (Germany) and IGN (Spain)}
\thanks{The integrated emission maps are available in electronic form
at the CDS via anonymous ftp to cdsarc.u-strasbg.fr }
}


\author{S. Spezzano \and P. Caselli \and L. Bizzocchi \and B.~M. Giuliano \and V. Lattanzi}

 \institute{Max Planck Institute for Extraterrestrial Physics, Giessenbachstrasse 1, 85748 Garching, Germany}


\abstract {Prior to star formation, pre-stellar cores accumulate matter towards the centre. As a consequence, their central density increases while the temperature decreases. Understanding the evolution of the chemistry and physics in this early phase is crucial to study the processes governing the formation of a star.} {We aim at studying the chemical differentiation of a prototypical pre-stellar core, L1544, by detailed molecular maps. In contrast with single pointing observations, we performed a deep study on the dependencies of chemistry on physical and external conditions.} 
{We present the emission maps of 39 different molecular transitions belonging to 22 different molecules in the central 6.25 arcmin$^2$ of L1544. We classified our sample in five families, depending on the location of their emission peaks within the core. Furthermore, to systematically study the correlations among different molecules, we have performed the principal component analysis (PCA) on the integrated emission maps. The PCA allows us to reduce the amount of variables in our dataset. Finally, we compare the maps of the first three principal components with the H$_2$ column density map, and the T$_{dust}$ map of the core.} 
{The results of our qualitative analysis is the classification of the molecules in our dataset in the following groups: (i) the $c$-C$_3$H$_2$ family (carbon chain molecules like C$_3$H and CCS), (ii) the dust peak family (nitrogen-bearing species like N$_2$H$^+$), (iii) the methanol peak family (oxygen-bearing molecules like methanol, SO and SO$_2$), (iv) the HNCO peak family (HNCO, propyne and its deuterated isotopologues). Only HC$^{18}$O$^+$ and $^{13}$CS do not belong to any of the above mentioned groups. The principal component maps allow us to confirm the (anti-)correlations among different families that were described in a first qualitative analysis, but also points out the correlation that could not be inferred before. For example, the molecules belonging to the dust peak and the HNCO peak families correlate in the third principal component map, hinting on a chemical/physical correlation.}
{The principal component analysis has shown to be a powerful tool to retrieve information about the correlation of different molecular species in L1544, and their dependence on physical parameters previously studied in the core. }

   \maketitle

\section{Introduction}
Starless cores represent the earliest phase of star formation. As at this stage the protostar is not yet formed, it is possible to study the chemistry and physics taking place away from the complexity arising from the proto-stellar feedback.
Pre-stellar cores are starless cores on the verge of star formation, the central density of H$_2$ is higher than 10$^5$ cm$^{-3}$ and therefore they are thermally supercritical \citep{ket08}. L1544 is a well-studied pre-stellar core in Taurus. It is centrally concentrated \citep{war99}, relatively massive \citep{taf98}, it shows very high deuteration \citep{cra05}, and radiative transfer studies show that it is contracting in a quasi-static fashion \citep{ket08, ket10, ket15}.
While several single pointing observations have been performed to study the chemistry of L1544, fewer maps are available. N$_2$H$^+$ and CCS are the first molecules that have been mapped towards L1554 \citep{wil99, oha99}, and already hinted on a chemical inhomogeneity of the core. CCS is less abundant towards the dust peak, while N$_2$H$^+$ suffers less the depletion caused by the high density and low temperature. 
The lack of substantial depletion for N$_2$H$^+$ has been confirmed by several studies: \cite{cas02} for instance show that N$_2$D$^+$, even better than N$_2$H$^+$, traces the inner region of L1544, close to the dust peak \citep{war99}. \cite{ber02} found hints of depletion of N$_2$H$^+$ towards the centre of B68. 
The molecular differentiation in L1544 and other starless cores was furthermore studied by \cite{taf02} by mapping CO and CS isotopologues, NH$_3$ and N$_2$H$^+$. All the cores, including L1544,  show an abundance drop of CO and CS at the dust peak, a constant N$_2$H$^+$ abundance and an increase of NH$_3$ towards the centre. While the behaviour of CO, CS and N$_2$H$^+$ are explained by depletion models, the increase of ammonia towards the centre is not yet well understood (Caselli et al., submitted).
The maps of DNC and HN$^{13}$C in L1544 and other starless cores show that, unlike the ratio $N$[DCO$^+$]/$N$[HCO$^+$], the column density ratio $N$[DNC]/$N$[HNC] is not sensitive to the depletion factor \citep{hir03}. The peculiar absence of depletion of nitrogen-bearing species has been used by \cite{cra07} to study the thermal structure of the denser parts in L1544 by mapping NH$_3$ and NH$_2$D. 
Recently it has been pointed out that the chemical differentiation in L1544 is not limited to nitrogen- and carbon-bearing species. Carbon-bearing molecules further differentiate, and in particular methanol (CH$_3$OH) and cyclopropenylidene ($c$-C$_3$H$_2$) present a complementary morphology \citep{spe16}.

This paper is an extensive study on the chemical differentiation in L1544. Maps of over 30 species are presented, and their different spatial distributions are discussed. The analysis will be carried out using the principal component analysis (PCA). This method has been successfully applied in other multi-line studies of other regions of our Galaxy, such as Orion \citep{ung97, gra17} and supernova remnants \citep{neu07}. In \cite{gra17}, for example, the PCA has been used to study the correlation between the emission maps of 12 molecular lines in the south-western edge of Orion B giant molecular cloud over a field of view of 1.5 square degrees.

The paper is structured as follows: the observations are described in Section \ref{Observations}, a first qualitative analysis of the data is presented in Section \ref{results}, and the results of the multivariate analysis are described in Section \ref{analysis}. Section \ref{conclusions} summarises our results and conclusions.

\begin{table*}
\caption{Spectroscopic parameters of the observed lines, divided depending on the position of their emission peaks.}
\label{table:parameters}
\scalebox{0.9}{
\begin{tabular}{lcr@{.}lr@{.}lr@{.}lc}
\hline\hline
Molecule & Transition &  \multicolumn{2}{c}{Rest frequency} &  \multicolumn{2}{c}{E$_{up}$}&  \multicolumn{2}{c}{A} &rms\tablefootmark{a} \\
&    & \multicolumn{2}{c}{(MHz)}   & \multicolumn{2}{c}{(K)}  &  \multicolumn{2}{c}{($\times$10$^{-5}$ s$^{-1}$)} & (mK)\\
\hline
\textbf{$c$-C$_3$H$_2$ peak}& \multicolumn{2}{c}{}& \multicolumn{2}{c}{}& \multicolumn{2}{c}{}&\\
$c$-$^{13}$C$_3$H$_2$\tablefootmark{*} &  $J_{K_a,K_c}$ = 2$_{1,2}$ - 1$_{0,1}$ & 84185&634 & 6&30 & 2&17& 6 \\
$c$-C$_3$H$_2$\tablefootmark{*}   & $J_{K_a,K_c}$ = $3_{2,2}$ - 3$_{1,3}$    &  84727&688     &  16&10     &  1&04& 21\\
$c$-C$_3$H$_2$  & $J_{K_a,K_c}$ = $2_{0,2}$ - 1$_{1,1}$    &  82093&544     &  6&40    &  1&89 & 17\\
H$_2$CCC  &  $J_{K_a,K_c}$ = 5$_{1,5}$ - 4$_{1,4}$   &  102992&379 & 28&19 &  9&32 & 16 \\
H$_2$CCC\tablefootmark{*}   &  $J_{K_a,K_c}$ = 4$_{1,3}$ - 3$_{1,2}$   & 83933&699 & 23&43 &  4&82 & 13\\
C$_3$H\tablefootmark{*} ($^2$$\Pi_{1/2}$ $\Lambda=b$\tablefootmark{b})  & $J$ = 9/2 - 7/2 $F$ = 5 - 4  & 97995&166 & 12&54 & 6&12 & 7\\
C$_3$H ($^2$$\Pi_{1/2}$ $\Lambda=b$\tablefootmark{b})&$J$ = 9/2 - 7/2 $F$ = 4 - 3  & 97995&913 & 12&54 & 5&95 & 7\\
C$_4$H\tablefootmark{*}   &  $N$ = 12 - 11 $J$ = 25/2 - 23/2 $F$ = 12 - 11 $\&$ 13 - 12   & 114182&510  &  35&62 &0&63  & 10 \\
H$_2$CCO\tablefootmark{*}  &   $J_{K_a,K_c}$ = 5$_{1,5}$ - 4$_{1,4}$  &  100094&514  &  27&46 & 1&03 & 7\\
H$_2$CCO &   $J_{K_a,K_c}$ = 5$_{1,4}$ - 4$_{1,3}$  &  101981&429  &  27&73 & 1&09& 18 \\
HCCNC\tablefootmark{*}   & $J$ = 9 - 8  &  89419&300 &  21&46  &  3&38 & 16 \\
H$_2$CS\tablefootmark{*}   &  $J_{K_a,K_c}$ = 3$_{0,3}$ - 2$_{0,2}$  &  103040&452  &  9&89  &  1&48 & 16 \\
HCS$^+$\tablefootmark{*}   &  $J$ = 2 - 1  & 85347&890  & 6&14  & 0&11 & 15\\
C$^{34}$S\tablefootmark{*}   &  $J$ = 2 - 1 &  96412&949  & 6&94  &  1&60  & 32\\
CCS& $N, J$ = 8, 7 - 7, 6   &  99866&521  & 28&14 &4&40 & 11 \\
CCS&$N, J$ = 7, 6 - 6, 5&86181&391&23&34& 2&78& 14\\
CCS&$N, J$ = 7, 7 - 6, 6&90686&381&26&12& 3&29&15\\ 
CCS\tablefootmark{*} &$N, J$ =  8, 9 - 7, 8&106347&726&25&00& 5&48& 16\\ 
CH$_3$CN\tablefootmark{*}  & $J_K$ = 6$_0$ - 5$_0$  &110383&500&18&54&11&11 &17\\
HCC$^{13}$CN\tablefootmark{*}  &  $J$= 10 - 9 &90601&777 &23&92 &5&74 & 10 \\
\hline
\textbf{Dust peak}&\multicolumn{2}{c}{}& \multicolumn{2}{c}{}& \multicolumn{2}{c}{}&\\
$^{13}$CN\tablefootmark{*}   & $N$ = 1 - 0 $F_1$ = 2 - 1 $F_2$ = 2 - 1 $F$ = 3 - 2 & 108780&201 & 5&25 & 1&05 & 4 \\
H$^{13}$CN\tablefootmark{*}  &  $J$= 1 - 0 $F$=  2 - 1 & 86340&168  & 4&14  &  2&25 & 20\\
N$_2$H$^+$\tablefootmark{*}   &  $J$= 1 - 0  $F_1$ = 0 - 1 $F$ = 1 - 2 &  93176&265  &  4&47 &2&01 & 20\\
\hline
\textbf{Methanol peak}&\multicolumn{2}{c}{}& \multicolumn{2}{c}{}& \multicolumn{2}{c}{}&\\
CH$_3$OH\tablefootmark{*}    &   $J_{K_a,K_c}$ = 2$_{1,2}$ - 1$_{1,1}$ ($E_2$) &   96739&362       &  12&53\tablefootmark{c}   & 0&26 & 37\\
CH$_3$OH   &   $J_{K_a,K_c}$ = 0$_{0,0}$ - 1$_{1,1}$ ($E_1$-$E_2$) &  108 893&963      &  13&12\tablefootmark{c}   & 1&47 & 9\\
SO\tablefootmark{*}   &   $N, J$ = 2, 2 - 1, 1 &  86093&950  & 19&31  & 0&52 & 27\\
SO  &   $N, J$ = 3, 2 - 2, 1 &  109252&220  & 21&05  & 1&08 & 23\\
$^{34}$SO\tablefootmark{*}   &   $N, J$ = 2, 3 - 1, 2  &97715&317 & 9&09  & 1&07 & 10\\
SO$_2$\tablefootmark{*}  &  $J_{K_a,K_c}$ = 3$_{1,3}$ - 2$_{0,2}$  & 104029&418  & 7&74 &1&01& 18 \\
OCS\tablefootmark{*}   & $J$=7 - 6 &   85139&103 & 16&34 & 0&17 & 14 \\
HCO\tablefootmark{*}  & $N_{K_a,K_c}$ = 1$_{0,1}$ - 0$_{0,0}$ $J$ = 3/2 - 1/2 $F$ = 2 - 1 & 86670&760 & 4&18& 0&47 & 14\\
\hline
\textbf{HNCO peak}&\multicolumn{2}{c}{}& \multicolumn{2}{c}{}& \multicolumn{2}{c}{}&\\
CH$_2$DCCH\tablefootmark{*}   &   $J_{K_a,K_c}$ = 6$_{0,6}$ - 5$_{0,5} $  &  97080&728  &  16&31  & 0&30 & 7\\
CH$_3$CCD\tablefootmark{*}   &    $J_K$ = 6$_1$ - 5$_1$  &  93454&331  &  22&92  & 0&26& 7\\
CH$_3$CCH\tablefootmark{*}   &   $J_K$ = 5$_0$ - 4$_0$  &  102546&024  &  17&30  & 0&20& 32\\
CH$_3$CCH  &   $J_K$ = 6$_1$ - 5$_1$  &  85457&300  &  24&45  & 0&35 & 20\\
HNCO\tablefootmark{*}   &     $J_{K_a,K_c}$ = 4$_{0,4}$ - 3$_{0,3}$  & 87925&237  & 10&55 & 0&88 & 20\\  
HNCO &     $J_{K_a,K_c}$ = 5$_{0,5}$ - 4$_{0,4}$  & 109905&749 & 15&82 & 1&75 &6\\  
\hline
\textbf{Other}&\multicolumn{2}{c}{}& \multicolumn{2}{c}{}& \multicolumn{2}{c}{}&\\
$^{13}$CS\tablefootmark{*}  &  $J$= 2 - 1 &  92494&308 &  6&66  &  1&41 & 18\\
HC$^{18}$O$^+$\tablefootmark{*}  &  $J$ = 1 - 0  & 85162&223  &  4&09  &  3&64 & 15\\
\hline
\end{tabular}
}
\tablefoot{
\tablefoottext{*}{Molecular transitions included in the principal component analysis;}
\tablefoottext{a}{Average rms of the map;}
\tablefoottext{b}{A transition between lambda doublets of lower energy is designated as an $a$ component and between upper doublets as a $b$ component (Brown et al. 1975);}
\tablefoottext{c}{Energy relative to the ground 0$_{0,0}$, $A$ rotational state.}
}
\end{table*}

\begin{figure*}
 \centering
 \includegraphics [width=1\textwidth]{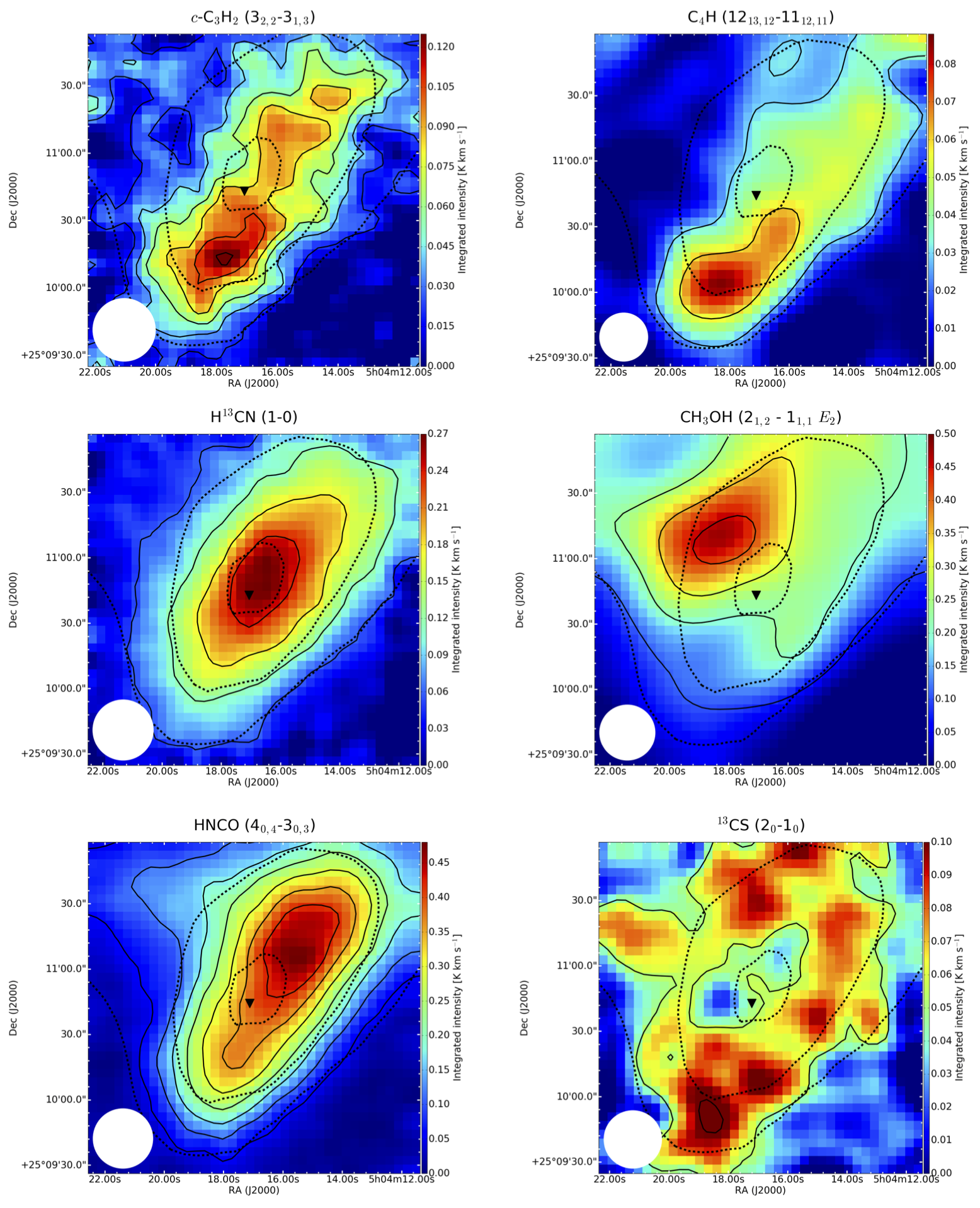}
 \caption{Sample of maps belonging to the different families observed towards L1544. The full dataset is shown in the Appendix A. The black dashed lines represent the 90$\%$, 50$\%$, and 30$\%$ of the H$_2$ column density peak value derived from Herschel maps \citep{spe16}, 2.8$\times$10$^{22}$ cm$^{-2}$. The solid lines represent contours of the molecular integrated emission starting with 3$\sigma$ with steps of 3$\sigma$ (the rms of each map is reported in Table \ref{table:parameters}). The dust peak \citep{war99} is indicated by the black triangle. The white circles represent the HPBW of the 30 m telescope.}
  \label{fig:sample}
\end{figure*}

\section{Observations}\label{Observations}
The emission maps towards L1544 have been obtained using the IRAM 30m telescope (Pico Veleta, Spain) in three different observing runs in October 2013, July 2015, and January 2017, and are shown in Figures \ref{fig:maps}-\ref{fig:maps4}. A sample of the observed maps is presented in Figure \ref{fig:sample}.
We performed a 2.5\arcmin $\times$2.5\arcmin on-the-fly (OTF) map centred on the source dust emission peak ($\alpha _{2000}$ = 05$^h$04$^m$17$^s$.21,  $\delta _{2000}$ = +25$^\circ$10$'$42$''$.8). Position switching was used, with the reference position set at (-180\arcsec,180\arcsec) offset with respect to the map centre. The observed transitions are summarised in Table \ref{table:parameters}. The EMIR E090 receiver was used with the Fourier Transform Spectrometer backend (FTS) with a spectral resolution of 50 kHz. The antenna moved along an orthogonal pattern of linear paths separated by 8$''$ intervals, corresponding to roughly one third of the beam FWHM. The mapping was carried out in good weather conditions ($\tau$ $\sim$ 0.03) and a typical system temperature of T$_{sys}$ $\sim$ 90-100 K. The data processing was done using the GILDAS software \citep{pet05} and CASA \citep{CASA}. All the emission maps presented in this paper have been gridded to a pixel size of 4$''$ with the CLASS software in the GILDAS package, this corresponds to 1/5-1/7 of the actual beam size, depending on the frequency. The integrated intensities have been calculated in the 6.7-7.7 km s$^{-1}$ velocity range.

\section{Results and Discussion}\label{results}
By making a qualitative comparison of their distributions in L1544, we can classify our sample of molecules in five different families: those with integrated intensity maps peaking nearby the (i) $c$-C$_3$H$_2$ peak ($c$-C$_3$H$_2$ and C$_4$H in Figure \ref{fig:sample}), (ii) dust peak (H$^{13}$CN in Figure \ref{fig:sample}), (iii) methanol peak (CH$_3$OH in Figure \ref{fig:sample}), (iv) HNCO peak (HNCO in Figure \ref{fig:sample}). Only a few do not belong to the above groups ($^{13}$CS in Figure \ref{fig:sample}). Figure \ref{fig:sample} shows some of the molecules presenting different spatial distribution within L1544, the whole observed sample is reported in Figures \ref{fig:maps}-\ref{fig:maps4}. 

\subsection{$c$-C$_3$H$_2$ peak}
The molecules belonging to this category peak towards the south-east of the core, see the maps in Figure \ref{fig:maps}. As already discussed in \cite{spe16}, this spot corresponds to a sharp edge in the H$_2$ column density map and hence it is more affected by the interstellar radiation field (ISRF). The carbon-chain chemistry (CCC) is enhanced and as a consequence, C-rich molecules have their emission peak in this region. Indeed the molecules that we find being abundant in the $c$-C$_3$H$_2$ peak are: H$_2$CCC, C$_3$H, C$_4$H, H$_2$CCO, HCCNC, H$_2$CS, HCS$^+$, C$^{34}$S, CCS, CH$_3$CN, HCC$^{13}$CN and of course $c$-C$_3$H$_2$ and its $^{13}$C isotopologue $c$-H$^{13}$CC$_2$H. We note that only sulfur molecules containing carbon are present here. This could be due to the presence of S$^+$, which can readily react with hydrocarbons \citep{smi88}.

\subsection{Dust peak}
The molecules that are more abundant towards the dust peak are nitrogen containing molecules: H$^{13}$CN, $^{13}$CN and N$_2$H$^+$, see Figure \ref{fig:maps1}.
It is long known that nitrogen-bearing molecules suffer less depletion towards the dust peak of starless cores with respect to other molecules, CO for example \citep{cas99}. This is evident from our maps as well. While all the other molecules, no matter where they peak, avoid the central denser region of L1544, nitrogen-bearing molecules are more abundant towards the dust peak. The explanation for such behaviour is not yet clear. Laboratory studies show that the desorption rates and sticking probabilities of N$_2$ and CO are very similar \citep{bis06}, but the observations in L1544 and other starless cores show that CO depletes at higher temperatures and lower densities than N$_2$H$^+$. A possible explanation is due to the fact that CO is at least partially destroying N$_2$H$^+$ (and producing HCO$^+$), so that N$_2$H$^+$ thrives when CO starts to freeze out (e.g. \citealt{aik01}). Moreover, N$_2$ formation may still be ongoing in the gas phase when CO becomes heavily depleted (e.g. \citealt{flo06}).

\subsection{Methanol peak}
\cite{biz14} presented the first map of methanol towards L1544, showing a clear peak towards the north of the dust continuum peak. We have recently pointed out that the methanol distribution differs from that of cyclopropenylidene, $c$-C$_3$H$_2$ \citep{spe16}. This leads to a spatial differentiation within C-bearing molecules not yet known. Here we present the maps of several molecules peaking at the methanol peak: SO, $^{34}$SO, SO$_2$, OCS and methanol, see Figure \ref{fig:maps2}. Three hyperfine components of HCO are also present in our data, but the achieved signal-to-noise ratio allows us to map only the  $N_{K_a,K_c}$ = 1$_{0,1}$ - 0$_{0,0}$ $J$ = 3/2 - 1/2 $F$ = 2 - 1 transition. Despite many of the molecules peaking in this position contain sulfur, we do not believe that sulfur chemistry is playing a crucial role in the chemical differentiation that we observe. In fact, the sulfur-bearing species that are carbon-based are more abundant towards the $c$-C$_3$H$_2$ peak. Sulfur-bearing molecules here all contain oxygen (unlike those toward the $c$-C$_3$H$_2$ peak).
\cite{jim16} show that O-bearing complex molecules (COM) like methyl formate and dimethyl ether are also more abundant towards the methanol peak while N-bearing COM like HCCNC, CH$_3$CN and CH$_2$CHCN are more abundant towards the dust peak. From our data we can confirm that O-bearing molecules are indeed more abundant in the low density shell towards the north of L1544, while HCCNC and CH$_3$CN peak towards the south-east like the other C-bearing molecules. The O-bearing molecules deplete towards the dust peak and are less abundant towards the south where the C/O atomic ratio is quite large, because of the photodissociation of CO, and the formation rate of carbon chains is very fast (e.g. \citealt{her89}).

\subsection{HNCO peak}
Four different species have their emission peak in none of the three positions discussed above. These species are CH$_3$CCH and its deuterated isotopologues, CH$_2$DCCH and CH$_3$CCD, and HNCO, see Figure \ref{fig:maps3}. The lines of CH$_3$CCH and HNCO are relatively bright and if their emission is optically thick, the distribution might be affected and hence not representative of the whole core. We should be able to estimate the optical depth based on the upper limit of the $^{13}$C isotopologue, assuming a $^{12}$C/$^{13}$C of 77 (Wilson $\&$ Rood 1994). Unfortunately we did not detect any lines of the $^{13}$C isotopologue of propyne, CH$_3$CCH, so we cannot deduce its spatial distribution from the $^{13}$C isotopologue. Instead, we have observed two deuterated isotopologues, CH$_2$DCCH and CH$_3$CCD, and they also peak towards the HNCO peak. Given the fact that we believe that it would be quite surprising for a deuterated isotopologue to be optically thick, an additional molecular peak is indeed present in L1544 towards the north-west, and not produced by optical depth effects. We call it here the HNCO peak.
We have also mapped two lines of HNCO, and they both show their emission peak towards the north-west. No $^{13}$C isotopologue was observed, so we cannot rule out that the emission peak shown by HNCO is caused by large optical depth by comparing with a rare isotopologue. We have instead used the online RADEX tool, a one-dimensional non-LTE radiative transfer code \citep{van07}. We have checked the optical depth of the two observed lines of HNCO by assuming a kinetic temperature ranging from 8 to 12 K and a density from 1$\times$10$^{5}$ to 1$\times$10$^{6}$ H$_2$ molecules per cubic centimetre and we obtain optical depths in the range of 0.1 to 0.3. As a consequence we can safely assume that the spatial distribution of HNCO is not affected by a large optical depth.

HNCO contains all the atoms present in prebiotic molecules, and its correlation with formamide (NH$_2$CHO) has been discussed in the literature \citep{lop15, bar15}. In case the HNCO peak was a genuine molecular peak in L1544, it would be the best place to look for prebiotic molecules like formamide, and shine some light on the formation of life-seeds in the earliest phases of star formation.

\subsection{Other}
Two molecules do not fall in neither of the four categories described above, see Figure \ref{fig:maps4}. HC$^{18}$O$^+$ is not peaking towards the $c$-C$_3$H$_2$ peak, but it is clearly depleted towards the centre of L1544. A similar behaviour was already observed by \cite{cas02} for H$^{13}$CO$^+$. 
$^{13}$CS shows a clear "donut" shape around L1544, indicating very clearly its depletion towards the dust peak, and peaking towards both the $c$-C$_3$H$_2$- and methanol peak. The map of the $J$ = 2 - 1 transition of CS, main isotopologue, was reported in \cite{taf02} and presents a diffuse and fragmented distribution compared to the N$_2$H$^+$ and NH$_3$ maps.  Both molecules (CS and HC$^{18}$O$^+$) are expected to be quite abundant in the molecular cloud within which L1544 is embedded, so their distribution may be affected by the large scale structure of the cloud.

\section{Multivariate analysis}\label{analysis}
\begin{figure*}[h]
 \centering
 \includegraphics [width=1\textwidth]{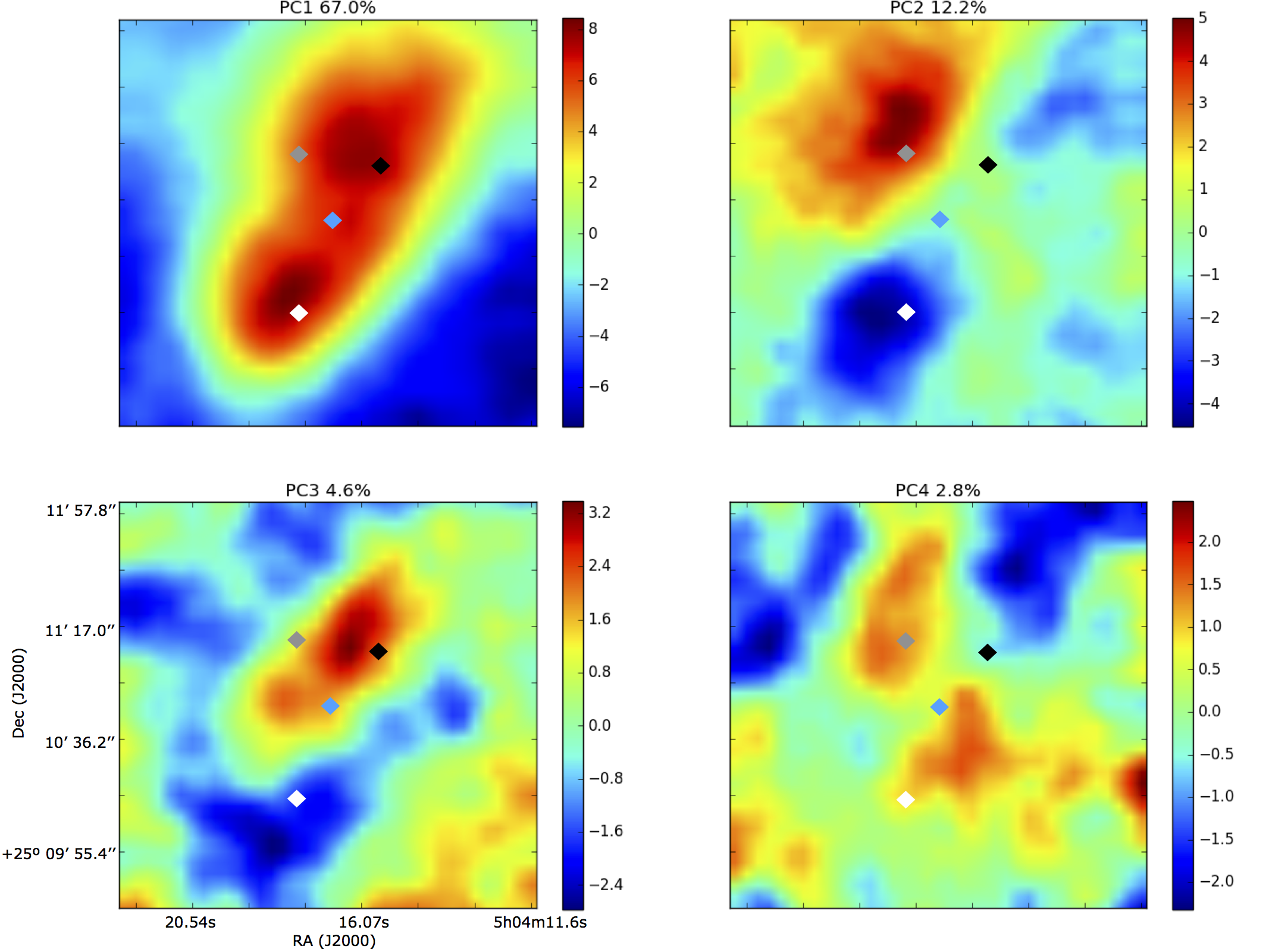}
 \caption{Maps of the first four principal components obtained by performing the PCA on the standardised data. The maps are constructed by summing for each pixel the contribution of each molecular transition scaled by the values reported in Table \ref{table:eigen}, i.e. they represent each pixel projected in the space of the principal components. The percentages represent the amount of correlation that can be reproduced by the single principal component. The blue, black, white and grey diamonds indicate the dust, the HNCO, the $c$-C$_3$H$_2$, and the methanol peaks respectively.}
  \label{fig:7}
\end{figure*}

\begin{figure}[h]
 \centering
 \includegraphics [width=0.5\textwidth]{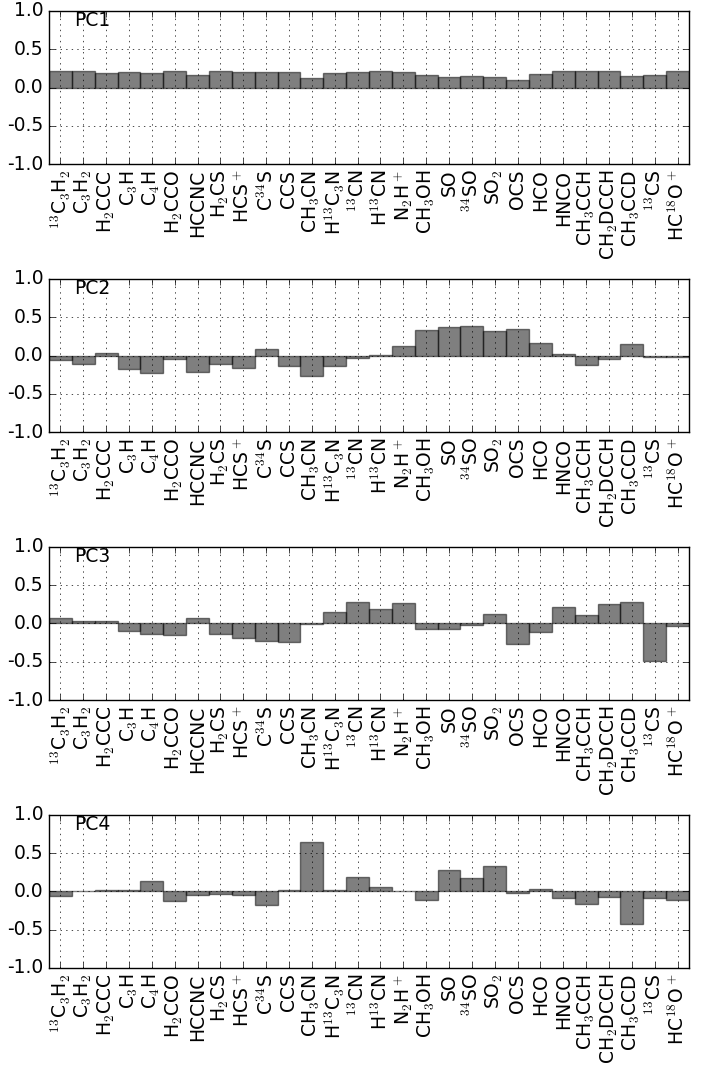}
 \caption{Contribution of each molecular transition to the first four PC, obtained by performing the PCA on the standardised data.}
  \label{fig:8}
\end{figure}

\begin{figure}[h]
 \centering
 \includegraphics [width=0.5\textwidth]{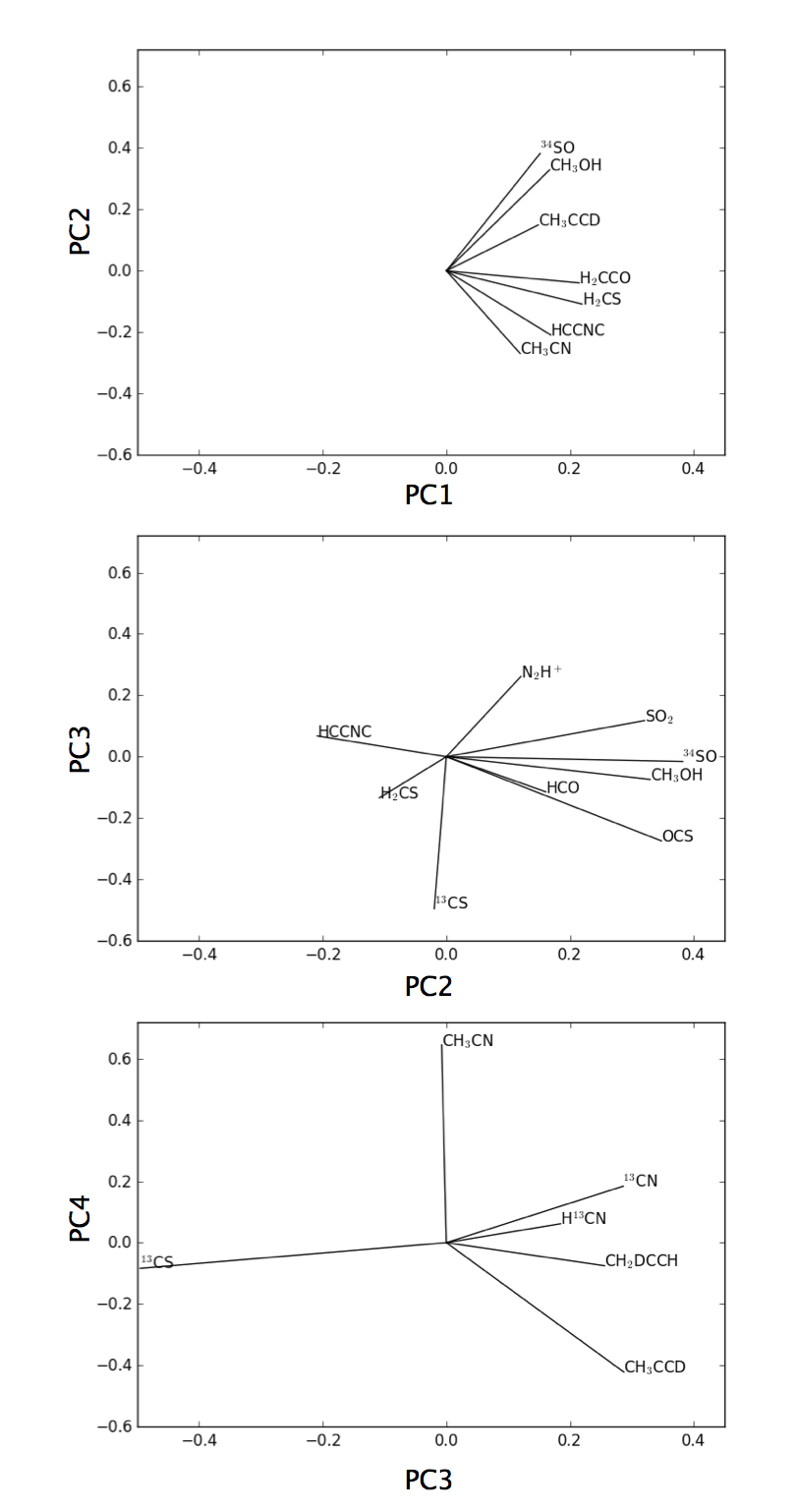}
 \caption{Correlation wheels where each molecule has as coordinates their correlation coefficients to each PC, obtained by performing the PCA on the standardised data.}
  \label{fig:9}
\end{figure}

To make an unbiased study of our large data set and systematically study the correlations among different molecules, we have performed the principal component analysis (PCA) on integrated emissions in L1544. We have used the PCA implementation available in the Python package "scikit-learn" \citep{ped11}.
The PCA is a multivariate analysis technique that allows us to reduce the dimensionality of the variables in our data set. It is widely used in natural sciences, and it has also been used to study the physics and chemistry of the ISM \citep{ung97, neu07, lo09, mel11, jon12, gra17}. 
The PCA method has been extensively described in \cite{ung97}. Briefly, the principal components (PCs) will be derived by finding the eigenvalues and eigenvectors of a covariance/correlation matrix, with the eigenvalues being the variance (or correlation) accounted for by each PC, and the eigenvectors the contributions of each molecule to the each PC.\\
When performing the PCA, we treat our data as 1369 points (37 $\times$ 37 pixels in each map), in a space with 28 dimensions (the number of molecular transitions that we use in the analysis). In our analysis we have included one molecular transition for each molecule that we have observed. In the case of molecules that have been mapped in several transitions, we have selected the map with the best signal-to-noise ratio. The molecular transitions included in the PCA are marked with a star in Table \ref{table:parameters}. In order to reduce the number of variables, we describe the variations in the 28 maps as a superimposition of fewer variables, the principal components. The maps of the PC will help us identify the regions within L1544 with stronger (anti-)correlations among the different molecules.
As done in the literature, we have standardised the data before performing the PCA, meaning that they are mean-centred and normalised. The standardised value is described as follows:\\
\begin{eqnarray*}
x_{std} &=& \frac{x-\mu}{\sigma},\\
\end{eqnarray*}

\noindent with $\mu$ being the mean value and $\sigma$ being the standard deviation for each map. The data before and after the standardisation are reported in the Appendix (Figure \ref{fig:10} and \ref{fig:12}), where a more detailed explanation on the need for preprocessing of the data is presented.  \\
The maps have not been convolved to a common angular resolution before performing the PCA. We have run some tests, and the effect of convolving all the maps to 29$''$ is negligible.
Table \ref{table:eigen} reports the results of our principal component analysis, i.e. the eigenvalues and eigenvectors of the first four principal components, and Figure \ref{fig:7} shows the projection of our data onto the PC, i.e. the principal component maps. It is mathematically possible to compute up to 28 independent PC, as many as the number of our maps. Figure \ref{fig:7} shows the maps of the first four principal components because for the purpose of our study it is enough to focus on those carrying most of the correlation, and the first four PC account for more than 85$\%$ of the correlation in our dataset. 
\\ Figure \ref{fig:8} shows the contribution of each emission line to each of the first four PC. The PC maps have no physical meaning, but they are very useful when compared with physical maps of L1544 to find possible (anti-)correlations among species and with the different physical conditions across the core. Figure \ref{fig:9} shows the eigenvectors of the first four PC plotted as correlation wheels. For the clarity of the plot, only a fraction of the molecules are shown, selected among the ones showing the most prominent (anti-)correlation features. \\

\begin{table}
\caption{Eingenvalues and eigenvectors derived from Principal Component Analysis on the standardised data.}
\label{table:eigen}
\scalebox{0.9}{
\begin{tabular}{l|r@{.}lr@{.}lr@{.}lr@{.}l}
\hline\hline
&  \multicolumn{2}{c}{PC1} &  \multicolumn{2}{c}{PC2} &  \multicolumn{2}{c}{PC3}&  \multicolumn{2}{c}{PC4} \\
\hline
Normalised eigenvalues & 0&67&0&12&0&05&0&03\\
\hline
$c$-$^{13}$C$_3$H$_2$ &0&22 & -0&06 & 0&07 & -0&06 \\
$c$-C$_3$H$_2$  &0&21&-0&11 & 0&03 & \multicolumn{2}{c}{4$\times$10$^{-3}$}\\
H$_2$CCC  &0&19  & 0&04 &0&03&0&01 \\
C$_3$H  &0&20 & -0&18 &-0&10 &0&02\\
C$_4$H  &0&19  &-0&22 &-0&13&0&13 \\
H$_2$CCO & 0&21&-0&04&-0&15&-0&12\\
HCCNC  &0&17 &-0&21&0&07&-0&05 \\
H$_2$CS  &0&22&-0&11&-0&13&-0&04 \\
HCS$^+$  &  0&20&-0&16&-0&19&-0&04\\
C$^{34}$S  &0&20 & 0&08 &-0&23 &-0&18\\
CCS&0&21&-0&14&-0&25&0&02\\
CH$_3$CN & 0&12&-0&27&\multicolumn{2}{c}{7$\times$10$^{-3}$}&0&64\\
HCC$^{13}$CN & 0&19 &-0&14&0&14&\multicolumn{2}{c}{9$\times$10$^{-3}$} \\
$^{13}$CN  &0&20 &-0&03 & 0&29 &0&18\\
H$^{13}$CN &  0&22&\multicolumn{2}{c}{2$\times$10$^{-3}$}&0&18&0&06\\
N$_2$H$^+$  &0&21&0&12&0&26&\multicolumn{2}{c}{9$\times$10$^{-3}$}\\
CH$_3$OH   &0&17  &0&33&-0&07&-0&11\\
SO  &0&14&0&37&-0&07&0&28\\
$^{34}$SO  &0&15 &0&38& -0&02 &0&17\\
SO$_2$ &0&14 &0&32&0&12&0&33 \\
OCS  &0&10&0&35&-0&27& -0&01 \\
HCO & 0&18&0&16&-0&11&0&03\\
CH$_2$DCCH  &0&21 &-0&04&0&26&-0&08\\
CH$_3$CCD  &  0&15 &0&15&0&29&-0&42\\
CH$_3$CCH  &0&21&-0&12&0&11&-0&17\\
HNCO  & 0&22&0&02&0&21&-0&08\\  
$^{13}$CS  &0&16& -0&02&-0&50&-0&08\\
HC$^{18}$O$^+$ & 0&21&-0&01&-0&04&-0&10\\
\hline
\end{tabular}
}
\end{table}

The first PC positively correlates with all of the 28 molecular transitions observed, see Table \ref{table:eigen}. Its map, see Figure \ref{fig:7}, looks like the shape of L1544 as seen in the continuum with the difference that while the latter peaks at the denser part of L1544 (the dust peak shown as a black triangle in the maps in Figures \ref{fig:maps}-\ref{fig:maps4}), the PC1 map peaks at the $c$-C$_3$H$_2$ and HNCO peaks. 
The first PC represents a weighted mean of all lines, in fact the distribution of most of the molecules in our sample (80$\%$) follows the density distribution described by the continuum map. Such a feature has been observed in all previous studies of the ISM with the PCA, e.g \cite{ung97}. 
The only exception are the molecules peaking towards the methanol peak, which are just 6 in a sample of 28, i.e. 20$\%$. In the PC1, the dust peak is not as intense as the HNCO peak, even if they both contribute with four molecular transitions to the dataset, because all of the molecules peaking in the $c$-C$_3$H$_2$ and HNCO peak are depleted towards the centre of the core. \\

The second PC map clearly represents the two most prominent molecular peaks in L1544, the $c$-C$_3$H$_2$ peak and the methanol peak. It correlates positively with all of the molecules peaking at the methanol peak, namely SO, SO$_2$, OCS, HCO, and methanol. On the other hand, all the C-chain molecules anti-correlate with the second PC. This differentiation has been already explained in \cite{spe16}, and it is related to a non-uniform exposure to the ISRF. Towards the South, where the C-chain molecules peak, L1544 presents a sharp edge in the H$_2$ column density, clearly shown in Figure 1 of \cite{spe16}. This region is less shielded from the ISRF and hence photochemistry is more prominent, keeping more carbon atoms available to support a rich carbon chemistry.
On the contrary, the methanol peak is farther away from the southern edge, hence more protected from the ISRF. Here carbon is mostly locked in CO. It is interesting to see that all of the molecules peaking towards the other two molecular peaks in L1544, the dust- and the HNCO peak, are not contributing substantially to the second PC.\\

While in the second PC map the difference between two molecular peaks is evident, and confirms our first qualitative classification, the third map brings together the dust and the HNCO peak. All of the molecules peaking in these two positions in fact correlate positively with the third PC. It is not trivial to see a chemical relationship among these molecules. While $^{13}$CN and N$_2$H$^+$ are both chemically related to N$_2$, it is less easy to find a common denominator with HNCO and CH$_3$CCH. A closer look into chemical models should help find the link between these four molecules.

By looking at the contributions to the fourth PC, see Figure \ref{fig:8}, we note that there is not a common behaviour for the different molecules belonging to the same family. The same happens also for the fifth PC, which is not shown in this paper. Given the amount of correlation that the fourth and fifth PC account for (less than 5$\%$), and the fact that we are aiming at using the PCA to find connections between the different families and physical quantities, we focus our attention only on the first three PC. 

\subsection{Effect of noise}
As done in \cite{gra17}, we have evaluated the effect of the noise on our results by performing the PCA on 5000 datasets which were constructed by adding random (gaussian) noise to the maps, according to the standard deviation of the noise in each single map. We have then computed the standard deviation of the PC coefficients for each molecule in the first 20 PC. We evaluate the effect of the noise by comparing the standard deviation of the PC coefficient for each molecule in each PC, with its mean within the 5000 resulting values ($\sigma$/$\mu$).
We find that the effect on the first three PC is completely negligible ($\sigma$/$\mu$ $<$ 5$\%$), and it is very small for the fourth and fifth PC ($\sigma$/$\mu$ $<$ 10$\%$). From the sixth PC on, the effect of the noise is instead relevant ($\sigma$/$\mu$ $\sim$ 30$\%$).

\subsection{Correlation with physical maps}
\begin{figure*}
 \centering
 \includegraphics [width=1\textwidth]{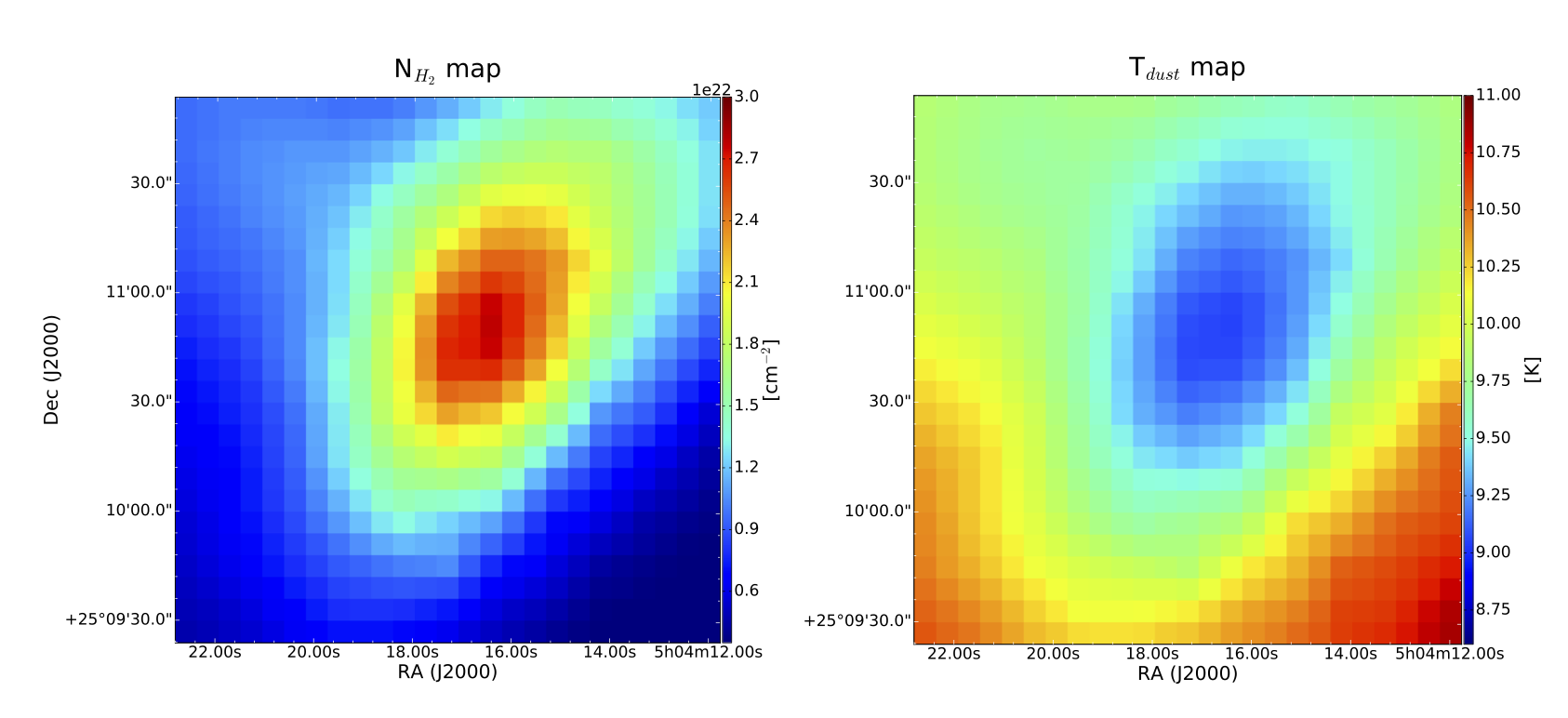}
 \caption{Maps of the H$_2$ column density and the dust temperature in L1544.}
  \label{fig:physical_maps}
\end{figure*}

As done in \cite{gra17}, we have computed the Spearman rank correlation coefficients between the first three PC maps and the physical maps of L1544. It is important to keep in mind that the physical maps that we present are correlated, while the PC maps are uncorrelated by definition. The physical maps that we have used are shown in Figure \ref{fig:physical_maps} and they are the H$_2$ column density map and the T$_{dust}$ map. The maps have been derived from Herschel/SPIRE data, following the method described in \cite{spe16} where the N$_{H_2}$ map has already been presented. The correlation coefficients are reported in Table \ref{table:spearman}. 

The N$_{H_2}$ map correlates very well with the map of the first principal component. This behaviour has been already reported in previous studies, e.g. \cite{gra17}, and it reflects the fact that the PC1 map can be interpreted like a global column density map. Given the fact that T$_{dust}$ anti-correlates with N$_{H_2}$ (high densities correspond to low temperatures), the first principal component shows a negative correlation with respect to them.

All the physical maps show a weak (anti-)correlation to the second principal component. The chemical differentiation described by the PC2 (for example methanol vs. $c$-C$_3$H$_2$) is in fact the result of large scale effects that are not taken into consideration in the small scale maps used for this analysis. This is very well explained by Figure 3 in \cite{spe16}, where the H$_2$ column density presents a steep drop towards the South-West, while the drop is more shallow towards the North-East. For this reason, the gas at the $c$-C$_3$H$_2$ peak in L1544 is less shielded from the UV illumination than the gas at the methanol peak. 

The third PC map, that well correlates with the molecules abundant at the dust and HNCO peak, does not show (anti-)correlation to any of the physical maps. Such a behaviour might be explained by the fact that these two classes of molecules present such spatial distribution because of a peculiar chemistry, or micro-physics properties (such as desorption rates and sticking probabilities) that are not affected by local variations in visual extinction (or N$_{H_2}$) and T$_{dust}$.

\begin{table}
\centering
\caption{Spearman rank correlation coefficients computed between the physical maps and the first three principal components }
\label{table:spearman}
\scalebox{1}{
\begin{tabular}{l|r@{.}lr@{.}lr@{.}l}
\hline\hline
&\multicolumn{2}{c}{PC1}& \multicolumn{2}{c}{PC2}&\multicolumn{2}{c}{PC3}\\
\hline
$N$(H$_2$)&0&96 &0&21 & -0&01 \\
T$_{dust}$ &-0&93& -0&29&0&04\\
\hline
\end{tabular}
}
\end{table}

\section{Conclusions}\label{conclusions}
We have presented the emission maps of 39 different molecular transitions towards the pre-stellar core L1544. The molecules present in our dataset have been divided in categories based on their spatial distribution. This qualitative approach has allowed us to recognise four molecular peaks in L1544: the $c$-C$_3$H$_2$ peak, the methanol peak, the dust peak and the HNCO peak.
The molecules belonging to the $c$-C$_3$H$_2$ family are carbon chain molecules: cyclic and linear C$_3$H$_2$, C$_3$H, C$_4$H, H$_2$CCO, H$_2$CS, C$^{34}$S, CCS, CH$_3$CN and HCC$^{13}$CN. The molecules belonging to the methanol peak family are characterised by the presence of Oxygen: methanol, SO, $^{34}$SO, SO$_2$, OCS, and HCO. Molecules chemically related to N$_2$ instead peak towards the dust peak: $^{13}$CN, H$^{13}$CN, and N$_2$H$^+$. The molecules belonging to the HNCO family are HNCO, CH$_3$CCH and its deuterated isotopologues. The chemical/physical connection among these molecules is still under investigation. $^{13}$CS and HC$^{18}$O$^+$ have a diffuse distribution within L1544, possibly affected by the large scale structure of the cloud.\\ In order to take into consideration underlying differences and similarities, we have performed the principal component analysis on a selected sample of molecular transitions in our dataset. The results of the PCA have confirmed the correlation among the molecules in the four categories depending on the spatial distribution, especially the dichotomy between the molecules peaking in the methanol and the $c$-C$_3$H$_2$ peak. The PCA has also shown correlation between the molecules in the HNCO and the dust peak of L1544, which could not be guessed just from a qualitative analysis. Further studies on the chemical link between these molecules will help shine some light on their correlation. \\
We calculated the Spearman rank correlation coefficients between the first three principal component maps and the maps of H$_2$ column density and the dust temperature in L1544. The first PC map, a weighted mean of all molecular transitions included in the analysis, correlates very well with the H$_2$ column density maps of L1544. The second PC, which reproduces well the contrast between methanol and $c$-C$_3$H$_2$, instead does not correlate substantially with the physical maps. This is in accordance with the large scale effects due to external illumination that we believe are responsible for the different spatial distribution of these two molecules. The third PC, which correlates with the molecules peaking both in the HNCO and dust peak, also shows no correlation with any of the physical maps, maybe suggesting that the link between these molecules has to be found in their chemistry or in the microphysics involved in the interaction with the ices.
A comparison with chemical modeling results will be presented in an upcoming paper (Spezzano, Sipil\"a et al., in prep.).\\
The very sensitive broadband receivers used currently in mm and sub-millimetre astronomy are delivering huge amount of data, and the use of multivariate analysis techniques, such as the PCA, might be of great help to get information on the links between physical and chemical conditions without using chemical modelling.

\noindent {\it Acknowledgement}\\
The authors wish to thank the anonymous referee for useful comments.
SS acknowledges the Christiane N\"usslein-Volhard Stiftung for financial support and S. Yazici for his help with the Python programming,
PC acknowledges the financial support of the European Research Council (ERC; project PALs 320620).



\begin{thebibliography}{}
\bibitem[Ag{\'u}ndez et al.(2008)]{agu08} Ag{\'u}ndez, M., Fonfr{\'{\i}}a, J.~P., Cernicharo, J., Pardo, J.~R., \& Gu{\'e}lin, M.\ 2008, \aap, 479, 493 

\bibitem[Ag{\'u}ndez et al.(2015)]{agu15} Ag{\'u}ndez, M., Cernicharo, J., \& Gu{\'e}lin, M.\ 2015, \aap, 577, L5 

\bibitem[Aikawa et al.(2001)]{aik01} Aikawa, Y., Ohashi, N., Inutsuka, S.-i., Herbst, E., \& Takakuwa, S.\ 2001, \apj, 552, 639 

\bibitem[Barone et al.(2015)]{bar15} Barone, V., Latouche, C., Skouteris, D., et al.\ 2015, \mnras, 453, L31 

\bibitem[Bachiller(1996)]{bac96} Bachiller, R.\ 1996, \araa, 34, 111 

\bibitem[Bergin et al.(2002)]{ber02} Bergin, E.~A., Alves, J., Huard, T., \& Lada, C.~J.\ 2002, \apjl, 570, L101 

\bibitem[Bisschop et al.(2006)]{bis06} Bisschop, S.~E., Fraser, H.~J., {\"O}berg, K.~I., van Dishoeck, E.~F., \& Schlemmer, S.\ 2006, \aap, 449, 1297 

\bibitem[Bizzocchi et al.(2014)]{biz14} Bizzocchi, L., Caselli, P., Spezzano, S., \& Leonardo, E.\ 2014, \aap, 569, A27 

\bibitem[Boogert et al.(2015)]{bog15} Boogert, A.~C.~A., Gerakines, P.~A., \& Whittet, D.~C.~B.\ 2015, \araa, 53, 541 

\bibitem[Buhl \& Snyder(1970)]{buh70} Buhl, D., \& Snyder, L.~E.\ 1970, \nat, 228, 267 

\bibitem[Caselli et al.(1995)]{cas95} Caselli, P., Myers, P.~C., \& Thaddeus, P.\ 1995, \apjl, 455, L77 

\bibitem[Caselli et al.(1999)]{cas99} Caselli, P., Walmsley, C.~M., Tafalla, M., Dore, L., \& Myers, P.~C.\ 1999, \apjl, 523, L165 

\bibitem[Caselli et al.(2002)]{cas02} Caselli, P., Walmsley, C.~M., Zucconi, A., et al.\ 2002, \apj, 565, 331 

\bibitem[Cernicharo et al.(1991)]{cer91} Cernicharo, J., Gottlieb, C.~A., Guelin, M., et al.\ 1991, \apjl, 368, L39 

\bibitem[Cernicharo et al.(2011)]{cer11} Cernicharo, J., Ag{\'u}ndez, M., Kahane, C., et al.\ 2011, \aap, 529, L3 


\bibitem[Coutens et al.(2016)]{cou16} Coutens, A., J{\o}rgensen, J.~K., van der Wiel, M.~H.~D., et al.\ 2016, \aap, 590, L6 

\bibitem[Crapsi et al.(2005)]{cra05} Crapsi, A., Caselli, P., Walmsley, C.~M., et al.\ 2005, \apj, 619, 379 

\bibitem[Crapsi et al.(2007)]{cra07} Crapsi, A., Caselli, P., Walmsley, M.~C., \& Tafalla, M.\ 2007, \aap, 470, 221 

\bibitem[Cuadrado et al.(2015)]{cua15} Cuadrado, S., Goicoechea, J.~R., Pilleri, P., et al.\ 2015, \aap, 575, A82 

\bibitem[Cummins et al.(1983)]{cum83} Cummins, S.~E., Green, S., Thaddeus, P., \& Linke, R.~A.\ 1983, \apj, 266, 331 

\bibitem[Dixon \& Woods(1977)]{dix77} Dixon, T.~A., \& Woods, R.~C.\ 1977, \jcp, 67, 3956 

\bibitem[Flower et al.(2006)]{flo06} Flower, D.~R., Pineau Des For{\^e}ts, G., \& Walmsley, C.~M.\ 2006, \aap, 449, 621 


\bibitem[Gottlieb et al.(1983)]{got83} Gottlieb, C.~A., Gottlieb, E.~W., Thaddeus, P., \& Kawamura, H.\ 1983, \apj, 275, 916 

\bibitem[Gottlieb et al.(1985)]{got85} Gottlieb, C.~A., Vrtilek, J.~M., Gottlieb, E.~W., Thaddeus, P., \& Hjalmarson, A.\ 1985, \apjl, 294, L55 

\bibitem[Gottlieb et al.(1986)]{got86} Gottlieb, C.~A., Gottlieb, E.~W., Thaddeus, P., \& Vrtilek, J.~M.\ 1986, \apj, 303, 446 

\bibitem[Gratier et al.(2017)]{gra17} Gratier, P., Bron, E., Gerin, M., et al.\ 2017, \aap, 599, A100 

\bibitem[Guarnieri et al.(1992)]{gua92} Guarnieri, A., Hinze, R., Kr{\"u}ger, M., et al.\ 1992, Journal of Molecular Spectroscopy, 156, 39 

\bibitem[Gudeman et al.(1981)]{gud81} Gudeman, C.~S., Haese, N.~N., Piltch, N.~D., \& Woods, R.~C.\ 1981, \apjl, 246, L47 

\bibitem[Guelin et al.(1982)]{gue82} Guelin, M., Friberg, P., \& Mezaoui, A.\ 1982, \aap, 109, 23 

\bibitem[Guelin et al.(1978)]{gue78} Guelin, M., Green, S., \& Thaddeus, P.\ 1978, \apjl, 224, L27 

\bibitem[Herbst \& Leung(1989)]{her89} Herbst, E., \& Leung, C.~M.\ 1989, \apjs, 69, 271 

\bibitem[Hirota et al.(2003)]{hir03} Hirota, T., Ikeda, M., \& Yamamoto, S.\ 2003, \apj, 594, 859 

\bibitem[Hollenbach et al.(1991)]{hol91} Hollenbach, D.~J., Takahashi, T., \& Tielens, A.~G.~G.~M.\ 1991, \apj, 377, 192 

\bibitem[Irvine et al.(1981)]{irv81} Irvine, W.~M., Hoglund, B., Friberg, P., Askne, J., \& Ellder, J.\ 1981, \apjl, 248, L113 

\bibitem[Irvine et al.(1989)]{irv89} Irvine, W.~M., Friberg, P., Kaifu, N., et al.\ 1989, \apj, 342, 871 

\bibitem[Jefferts et al.(1971)]{jef71} Jefferts, K.~B., Penzias, A.~A., Wilson, R.~W., \& Solomon, P.~M.\ 1971, \apjl, 168, L111 

\bibitem[Jim{\'e}nez-Escobar et al.(2014)]{jim14} Jim{\'e}nez-Escobar, A., Giuliano, B.~M., Mu{\~n}oz Caro, G.~M., Cernicharo, J., \& Marcelino, N.\ 2014, \apj, 788, 19 

\bibitem[Jim{\'e}nez-Serra et al.(2016)]{jim16} Jim{\'e}nez-Serra, I., Vasyunin, A.~I., Caselli, P., et al.\ 2016, \apjl, 830, L6 

\bibitem[Jones et al.(2012)]{jon12} Jones, P.~A., Burton, M.~G., Cunningham, M.~R., et al.\ 2012, \mnras, 419, 2961 

\bibitem[Johnson \& Strandberg(1952)]{joh52} Johnson, H.~R., \& Strandberg, M.~W.~P.\ 1952, \jcp, 20, 687 

\bibitem[Johnson et al.(1971)]{joh71} Johnson, D.~R., Powell, F.~X., \& Kirchhoff, W.~H.\ 1971, Journal of Molecular Spectroscopy, 39, 136 

\bibitem[Kaifu et al.(2004)]{kai04} Kaifu, N., Ohishi, M., Kawaguchi, K., et al.\ 2004, \pasj, 56, 69 

\bibitem[Kawaguchi et al.(1991)]{kaw91} Kawaguchi, K., Kaifu, N., Ohishi, M., et al.\ 1991, \pasj, 43, 607 

\bibitem[Kawaguchi et al.(1992)]{kaw92} Kawaguchi, K., Ohishi, M., Ishikawa, S.-I., \& Kaifu, N.\ 1992, \apjl, 386, L51 

\bibitem[Keto 
\& Caselli(2008)]{ket08} Keto, E., \& Caselli, P.\ 2008, \apj, 683, 238 

\bibitem[Keto 
\& Caselli(2010)]{ket10} Keto, E., \& Caselli, P.\ 2010, \mnras, 402, 1625 

\bibitem[Keto et al.(2015)]{ket15} Keto, E., Caselli, P., 
\& Rawlings, J.\ 2015, \mnras, 446, 3731 

\bibitem[Klemperer(1970)]{kle70} Klemperer, W.\ 1970, \nat, 227, 1230 

\bibitem[Liszt et al.(2012)]{lis12} Liszt, H., Sonnentrucker, P., Cordiner, M., \& Gerin, M.\ 2012, \apjl, 753, L28 

\bibitem[Liszt et al.(2014)]{lis14} Liszt, H.~S., Pety, J., Gerin, M., \& Lucas, R.\ 2014, \aap, 564, A64 

\bibitem[Lo et al.(2009)]{lo09} Lo, N., Cunningham, M.~R., Jones, P.~A., et al.\ 2009, \mnras, 395, 1021 

\bibitem[L{\'o}pez-Sepulcre et al.(2015)]{lop15} L{\'o}pez-Sepulcre, A., Jaber, A.~A., Mendoza, E., et al.\ 2015, \mnras, 449, 2438 

\bibitem[Lovas et al.(1992)]{lov92} Lovas, F.~J., Suenram, R.~D., Ogata, T., \& Yamamoto, S.\ 1992, \apj, 399, 325 

\bibitem[Madden et al.(1989)]{mad89} Madden, S.~C., Irvine, 
W.~M., Swade, D.~A., Matthews, H.~E., \& Friberg, P.\ 1989, \aj, 97, 1403 

\bibitem[Mallinson \& de Zafra(1978)]{mal78} Mallinson, P.~D., \& de Zafra, R.~L.\ 1978, Molecular Physics, 36, 827 

\bibitem[Marcelino et al.(2009)]{mar09} Marcelino, N., Cernicharo, J., Tercero, B., \& Roueff, E.\ 2009, \apjl, 690, L27 

\bibitem[Mart{\'{\i}}n et al.(2003)]{mar03} Mart{\'{\i}}n, S., Mauersberger, R., Mart{\'{\i}}n-Pintado, J., Garc{\'{\i}}a-Burillo, S., \& Henkel, C.\ 2003, \aap, 411, L465 

\bibitem[Melnick et al.(2011)]{mel11} Melnick, G.~J., Tolls, V., Snell, R.~L., et al.\ 2011, \apj, 727, 13 

\bibitem[McGuire et al.(2013)]{mcg13} McGuire, B.~A., Carroll, P.~B., Loomis, R.~A., et al.\ 2013, \apj, 774, 56 

\bibitem[McMullin et al.(2007)]{CASA} McMullin, J.~P., 
Waters, B., Schiebel, D., Young, W., 
\& Golap, K.\ 2007, Astronomical Data Analysis Software and Systems XVI, 376, 127 

\bibitem[M{\"u}ller et al.(2000)]{mul00} M{\"u}ller, H.~S.~P., Thorwirth, S., Bizzocchi, L., \& Winnewisser, G.\ 2000, Zeitschrift Naturforschung Teil A, 55,  

\bibitem[Nakajima et al.(2011)]{nak11} Nakajima, T., Takano, S., Kohno, K., \& Inoue, H.\ 2011, \apjl, 728, L38 

\bibitem[Neufeld et al.(2007)]{neu07} Neufeld, D.~A., Hollenbach, D.~J., Kaufman, M.~J., et al.\ 2007, \apj, 664, 890 

\bibitem[Nummelin et al.(1998)]{num98} Nummelin, A., Bergman, P., Hjalmarson, {\AA}., et al.\ 1998, \apjs, 117, 427 

\bibitem[Ohashi et al.(1999)]{oha99} Ohashi, N., Lee, S.~W., Wilner, D.~J., \& Hayashi, M.\ 1999, \apjl, 518, L41 

\bibitem[Ohishi et al.(1991)]{ohi91} Ohishi, M., Kawaguchi, K., Kaifu, N., et al.\ 1991, Atoms, Ions and Molecules: New Results in Spectral Line Astrophysics, 16, 387 

\bibitem[Pardo \& Cernicharo(2007)]{par07} Pardo, J.~R., \& Cernicharo, J.\ 2007, \apj, 654, 978 

\bibitem[Pedregosa et al.(2011)]{ped11} Pedregosa, F., Varoquaux, G., Gramfort, A., et al. \ 2011, Journal of Machine Learning Research, 12, 2825 

\bibitem[Penzias et al.(1974)]{pen74} Penzias, A.~A., Wilson, R.~W., \& Jefferts, K.~B.\ 1974, Physical Review Letters, 32, 701 

\bibitem[Pety(2005)]{pet05} Pety, J.\ 2005, SF2A-2005: 
Semaine de l'Astrophysique Francaise, 721 


\bibitem[Pratap et al.(1997)]{pra97} Pratap, P., Dickens, J.~E., Snell, R.~L., et al.\ 1997, \apj, 486, 862 

\bibitem[Saito et al.(1987)]{sai87} Saito, S., Kawaguchi, K., Yamamoto, S., et al.\ 1987, \apjl, 317, L115 

\bibitem[Sastry et al.(1981)]{sas81} Sastry, K.~V.~L.~N., Helminger, P., Charo, A., Herbst, E., \& De Lucia, F.~C.\ 1981, \apjl, 251, L119 

\bibitem[Sinclair et al.(1973)]{sin73} Sinclair, M.~W., Fourikis, N., Ribes, J.~C., et al.\ 1973, Australian Journal of Physics, 26, 85 

\bibitem[Sipil{\"a} et al.(2016)]{sip16} Sipil{\"a}, O., Spezzano, S., \& Caselli, P.\ 2016, \aap, 591, L1 

\bibitem[Smith et al.(1988)]{smi88} Smith, D., Adams, N.~G., Giles, K., \& Herbst, E.\ 1988, \aap, 200, 191 

\bibitem[Snyder \& Buhl(1973)]{sny73} Snyder, L.~E., \& Buhl, D.\ 1973, Nature Physical Science, 243, 45 

\bibitem[Snyder et al.(1976)]{sny76} Snyder, L.~E., Hollis, J.~M., \& Ulich, B.~L.\ 1976, \apjl, 208, L91 

\bibitem[Solomon et al.(1971)]{sol71} Solomon, P.~M., Jefferts, K.~B., Penzias, A.~A., \& Wilson, R.~W.\ 1971, \apjl, 168, L107 

\bibitem[Spezzano et al.(2016)]{spe16} Spezzano, S., Bizzocchi, L., Caselli, P., Harju, J., \& Br{\"u}nken, S.\ 2016b, \aap, 592, L11 

\bibitem[Spezzano et al.(2016b)]{spe16b} Spezzano, S., Gupta, H., Br{\"u}nken, S., et al.\ 2016, \aap, 586, A110 

\bibitem[Sutton et al.(1991)]{sut91} Sutton, E.~C., Jaminet, P.~A., Danchi, W.~C., \& Blake, G.~A.\ 1991, \apjs, 77, 255 

\bibitem[Tafalla et al.(1998)]{taf98} Tafalla, M., Mardones, 
D., Myers, P.~C., et al.\ 1998, \apj, 504, 900 

\bibitem[Tafalla et al.(2002)]{taf02} Tafalla, M., Myers, P.~C., Caselli, P., Walmsley, C.~M., \& Comito, C.\ 2002, \apj, 569, 815 

\bibitem[Tafalla et al.(2004)]{taf04} Tafalla, M., Myers, P.~C., Caselli, P., \& Walmsley, C.~M.\ 2004, \aap, 416, 191 

\bibitem[Thaddeus et al.(1981)]{tha81} Thaddeus, P., Guelin, M., \& Linke, R.~A.\ 1981, \apjl, 246, L41 

\bibitem[Thaddeus et al.(1985a)]{tha85} Thaddeus, P., Vrtilek, 
J.~M., \& Gottlieb, C.~A.\ 1985a, \apjl, 299, L63 

\bibitem[Thaddeus et al.(1985b)]{tha85b} Thaddeus, P., Gottlieb, C.~A., Hjalmarson, A., et al.\ 1985b, \apjl, 294, L49 


\bibitem[Turner(1977)]{tur77} Turner, B.~E.\ 1977, \apjl, 213, L75 

\bibitem[Turner et al.(1999)]{tur99} Turner, B.~E., Terzieva, R., \& Herbst, E.\ 1999, \apj, 518, 699 

\bibitem[Turner et al.(2000)]{tur00} Turner, B.~E., Herbst, E., \& Terzieva, R.\ 2000, \apjs, 126, 427 

\bibitem[Ungerechts et al.(1997)]{ung97} Ungerechts, H., Bergin, E.~A., Goldsmith, P.~F., et al.\ 1997, \apj, 482, 245 

\bibitem[van der Tak et al.(2007)]{van07} van der Tak, F.~F.~S., Black, J.~H., Sch{\"o}ier, F.~L., Jansen, D.~J., \& van Dishoeck, E.~F.\ 2007, \aap, 468, 627 

\bibitem[Vrtilek et al.(1990)]{vrt90} Vrtilek, J.~M., Gottlieb, .~A., Gottlieb, E.~W., Killian, T.~C., \& Thaddeus, P.\ 1990, \apjl, 364, L53 

\bibitem[Ward-Thompson et al.(1999)]{war99} Ward-Thompson, 
D., Motte, F., \& Andre, P.\ 1999, \mnras, 305, 143 

\bibitem[Williams et al.(1999)]{wil99} Williams, J.~P., Myers, P.~C., Wilner, D.~J., \& Di Francesco, J.\ 1999, \apjl, 513, L61 

\bibitem[Wilson \& Rood(1994)]{wil94} Wilson, T.~L., \& Rood, R.\ 1994, \araa, 32, 191 

\bibitem[Wyrowski et al.(2003)]{wyr03} Wyrowski, F., Schilke, P., Thorwirth, S., Menten, K.~M., \& Winnewisser, G.\ 2003, \apj, 586, 344 

\bibitem[Woods et al.(1975)]{woo75} Woods, R.~C., Dixon, T.~A., Saykally, R.~J., \& Szanto, P.~G.\ 1975, Physical Review Letters, 35, 1269 


\end{thebibliography}

\begin{appendix}

\section {Data set}\label{dataset}
We have mapped 39 different transitions of 22 different molecules and some of their isotopologues (Figures \ref{fig:maps}-\ref{fig:maps4}). This large data set allows us to study in detail the chemical differentiation across L1544. The spectroscopic parameters of the observed lines are summarised in Table \ref{table:parameters}. In this section we discuss the individual characteristics of the molecules that we have observed.\\

\begin{figure*}
 \centering
 \includegraphics [width=1\textwidth]{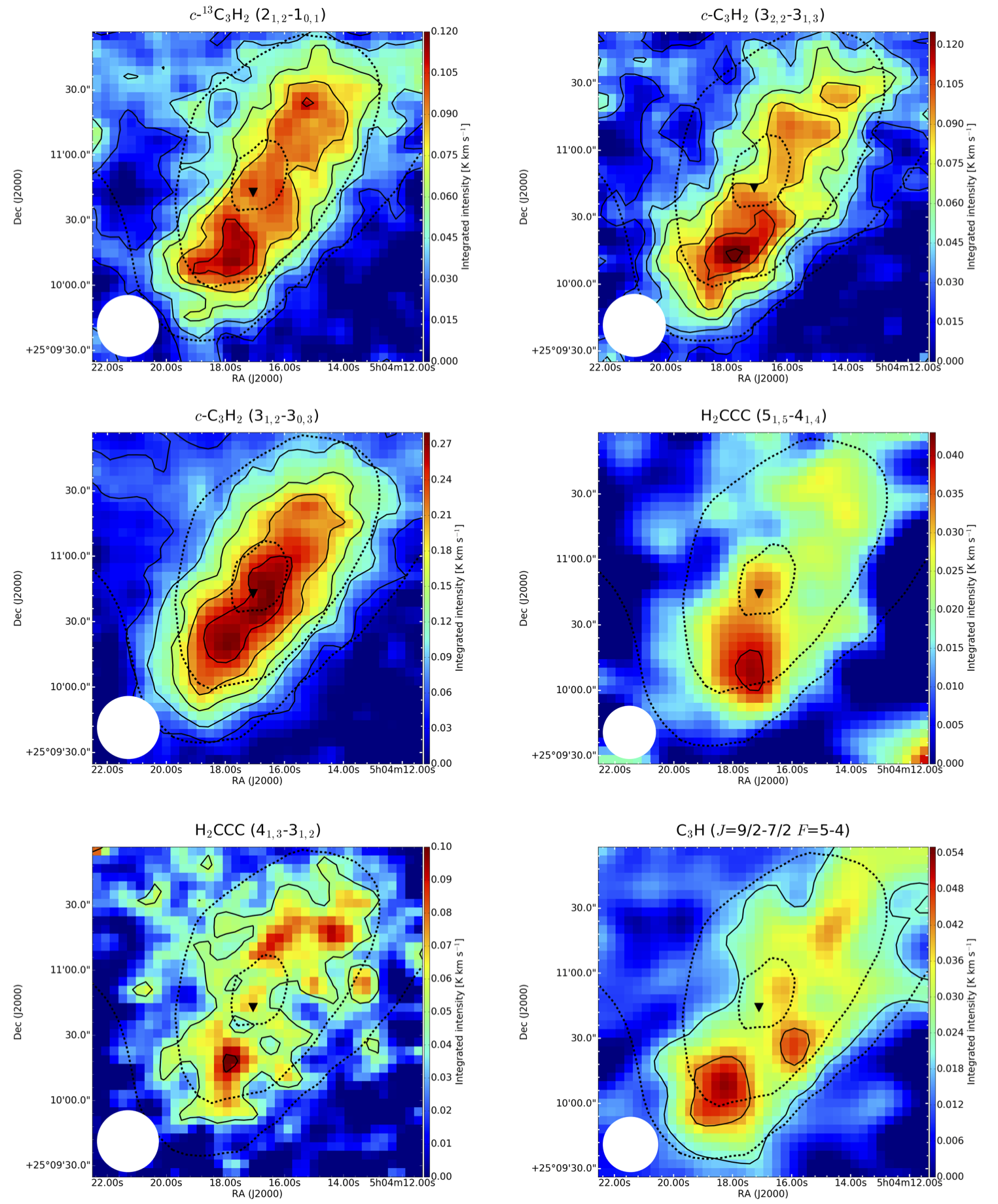}
 \caption{Maps of the molecules belonging to the $c$-C$_3$H$_2$ family observed in L1544. The black dashed lines represent the 90$\%$, 50$\%$, and 30$\%$ of the H$_2$ column density peak value derived from Herschel maps \citep{spe16}, 2.8$\times$10$^{22}$ cm$^{-2}$. The solid lines represent contours of the molecular integrated emission starting with 3$\sigma$ with steps of 3$\sigma$ (the rms of each map is reported in Table \ref{table:parameters}). The dust peak \citep{war99} is indicated by the black triangle. The white circles represent the HPBW of the 30 m telescope.}
  \label{fig:maps}
\end{figure*}


\begin{figure*}[h]
\ContinuedFloat
\captionsetup{list=off,format=cont}
\centering
 \includegraphics [width=1\textwidth]{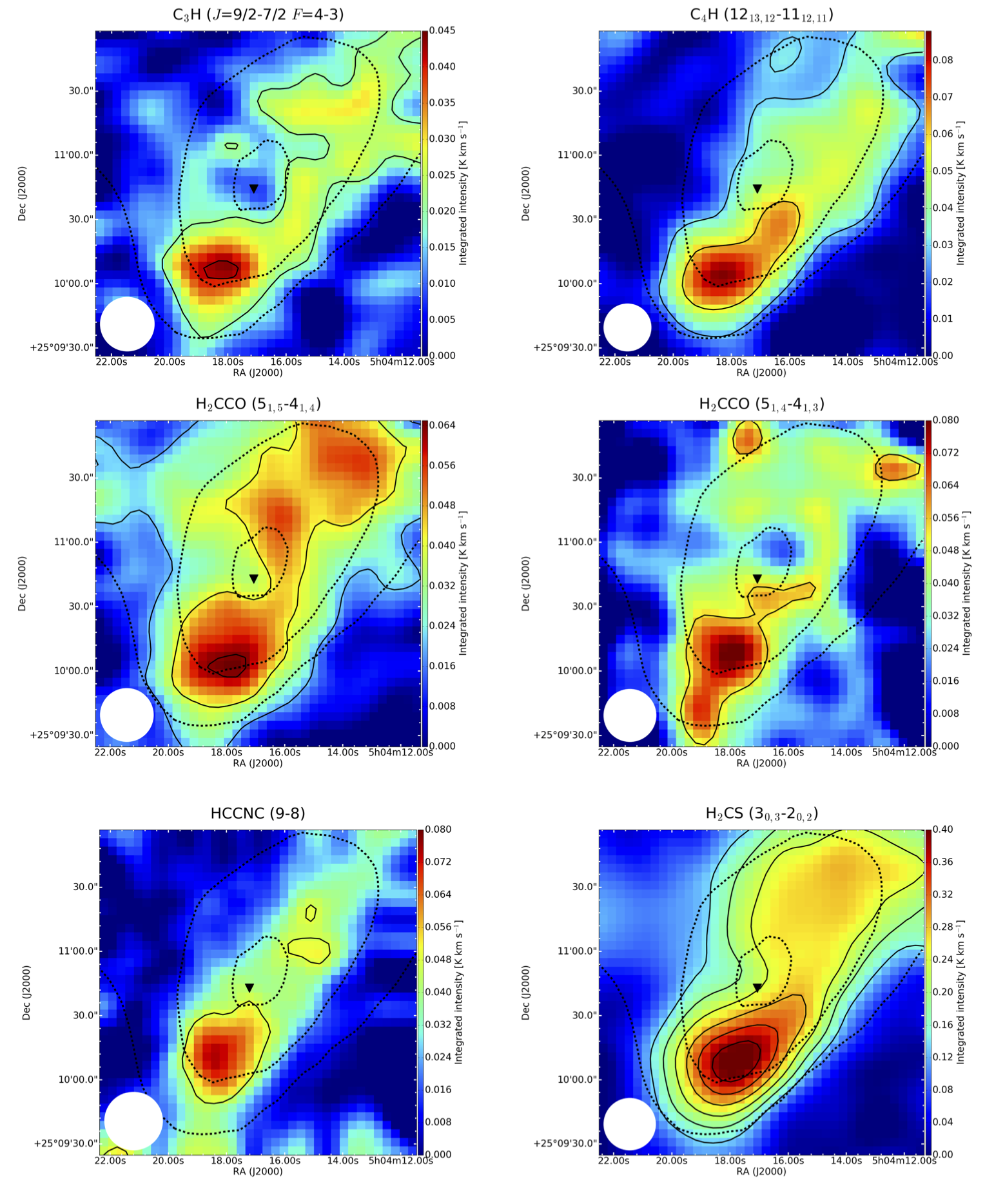}
 \caption{}
  \label{fig:}
\end{figure*}


\begin{figure*}[h]
\ContinuedFloat
\captionsetup{list=off,format=cont}
 \centering
 \includegraphics [width=1\textwidth]{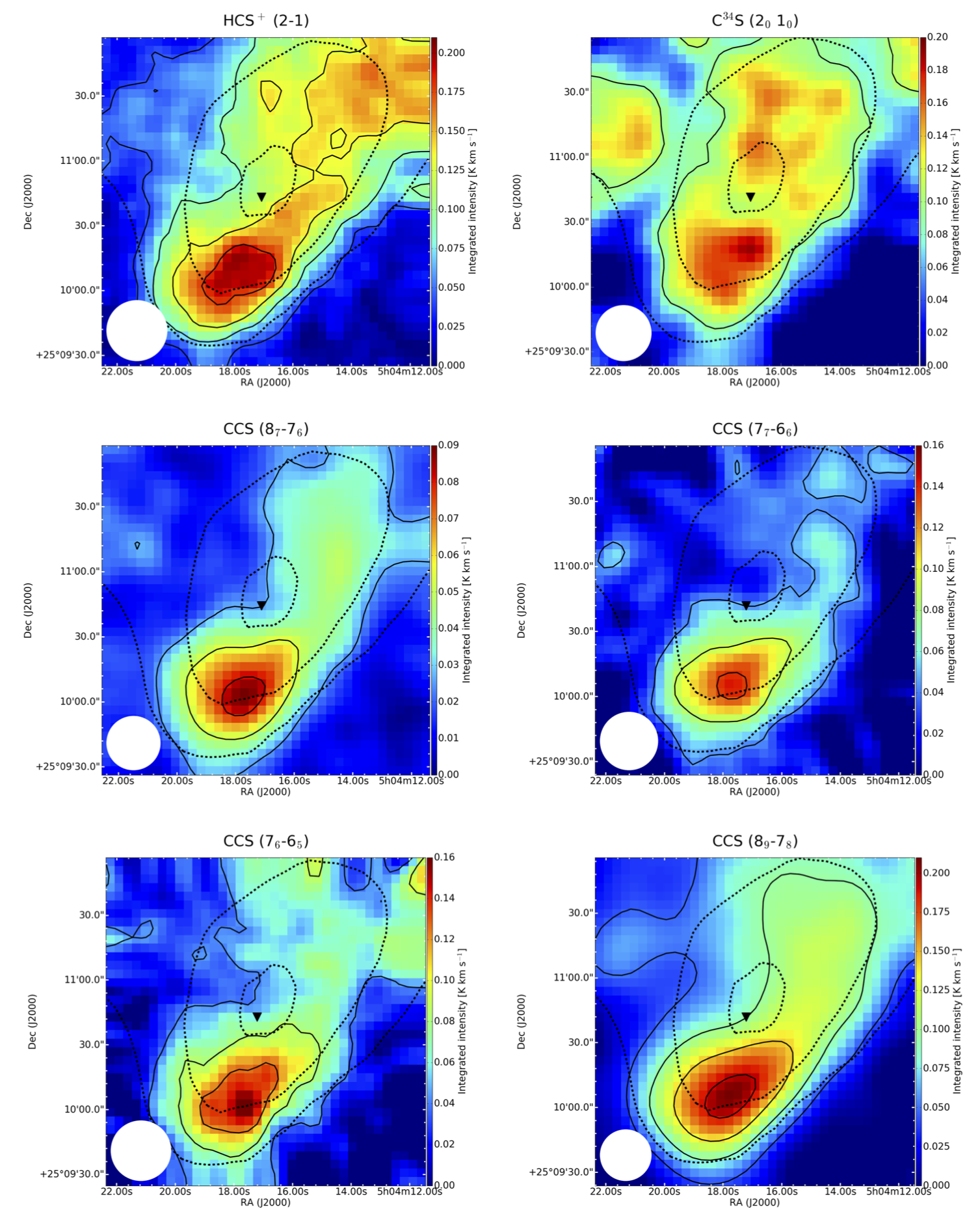}
  \caption{}
 \label{fig:}
\end{figure*}


\begin{figure*}[h]
\ContinuedFloat
\captionsetup{list=off,format=cont}
 \centering
 \includegraphics [width=1\textwidth]{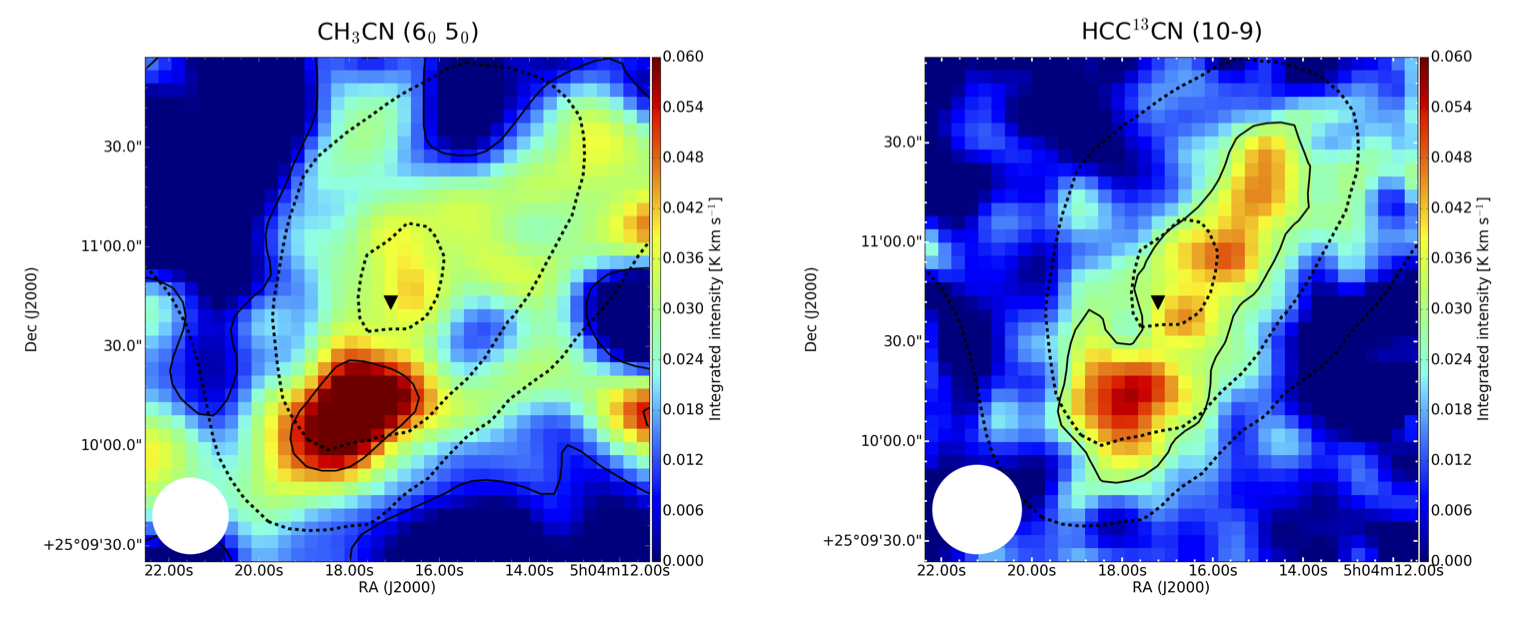}
  \caption{}
  \label{fig:}
\end{figure*}

\begin{figure*}[h]
 \centering
 \includegraphics [width=1\textwidth]{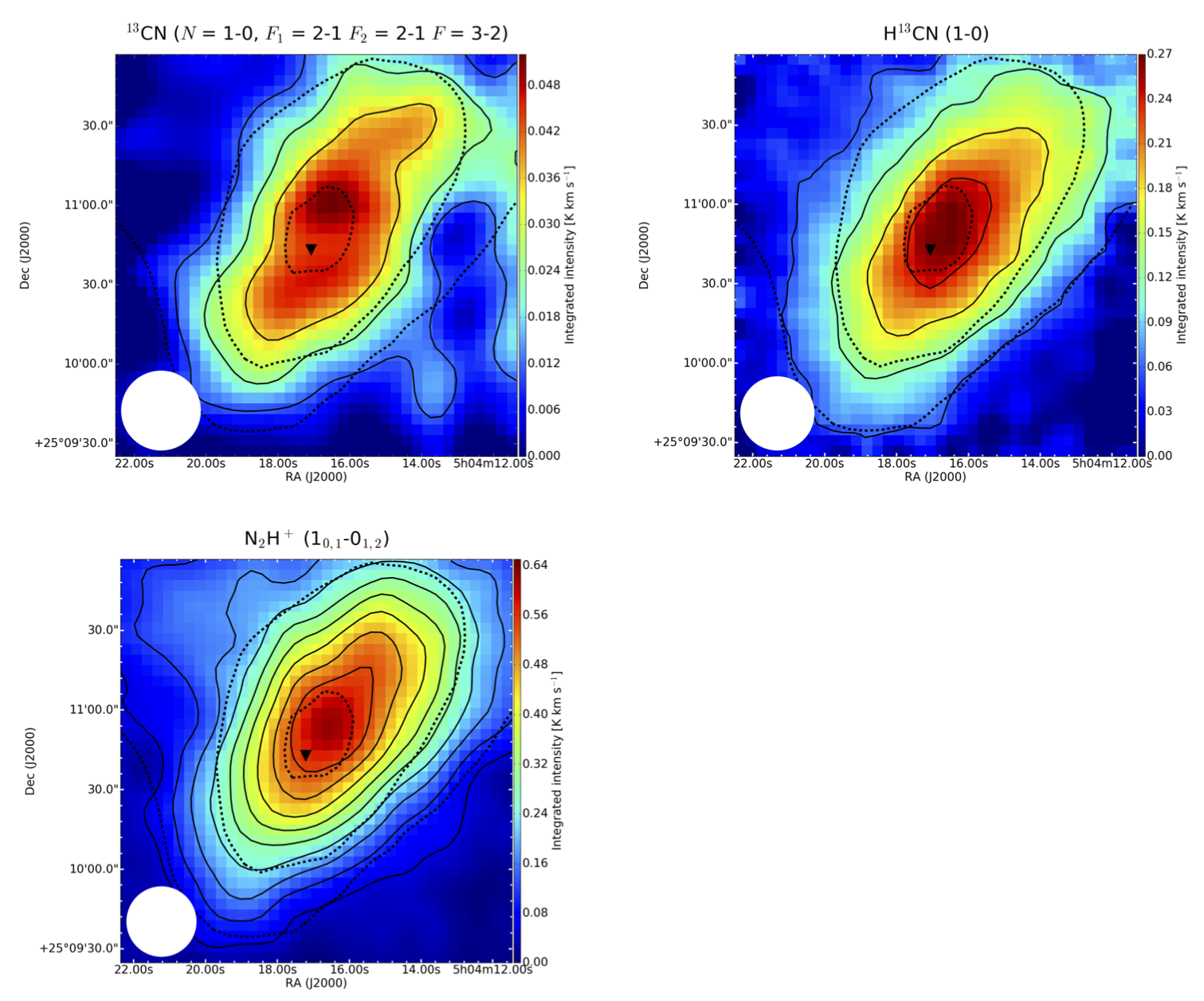}
  \caption{Maps of the molecules belonging to the dust peak family.}
  \label{fig:maps1}
\end{figure*}

\begin{figure*}[h]
 \centering
 \includegraphics [width=1\textwidth]{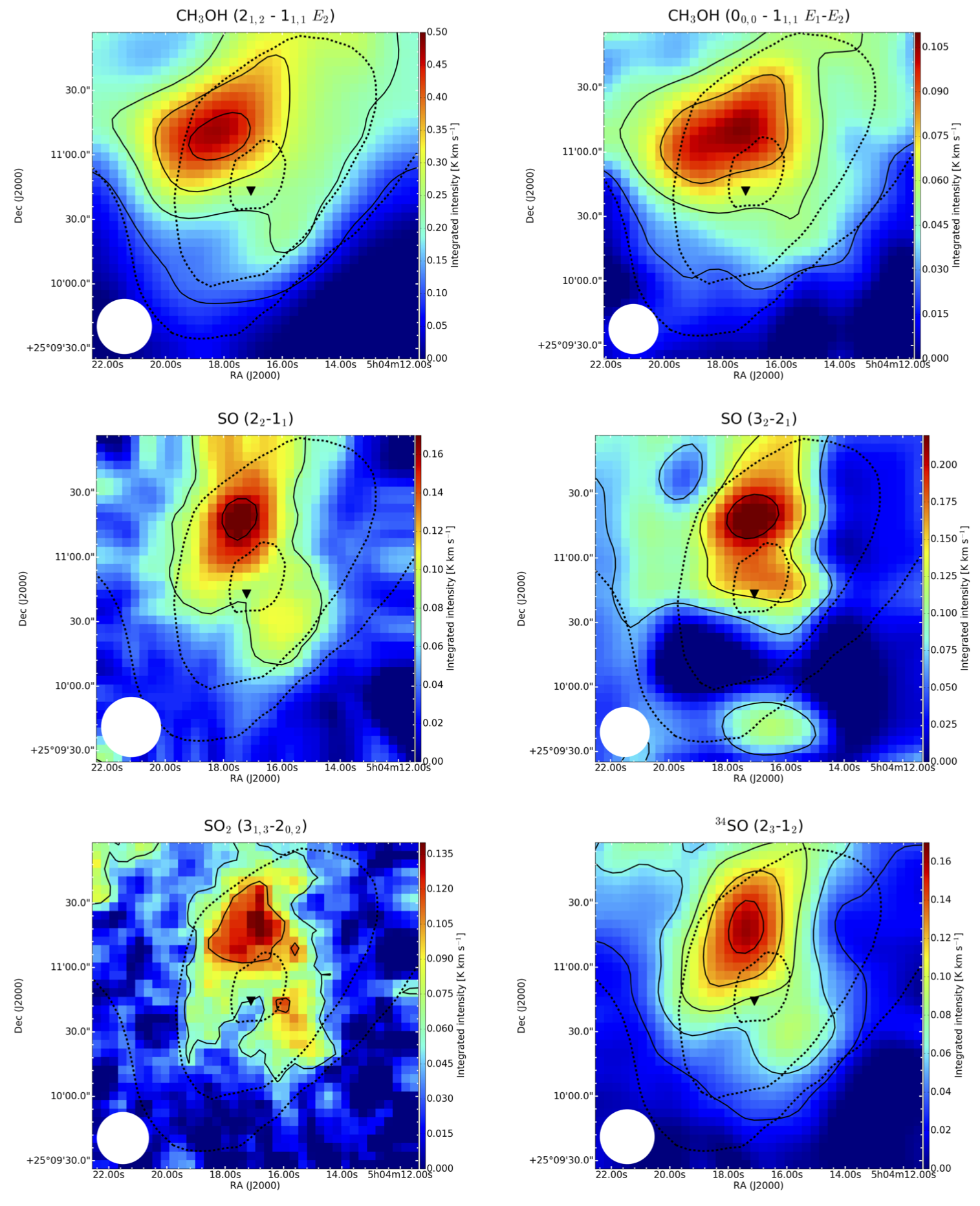}
  \caption{Maps of the molecules belonging to the methanol peak family.}
  \label{fig:maps2}
\end{figure*}


\begin{figure*}[h]
\ContinuedFloat
\captionsetup{list=off,format=cont}
 \centering
 \includegraphics [width=1\textwidth]{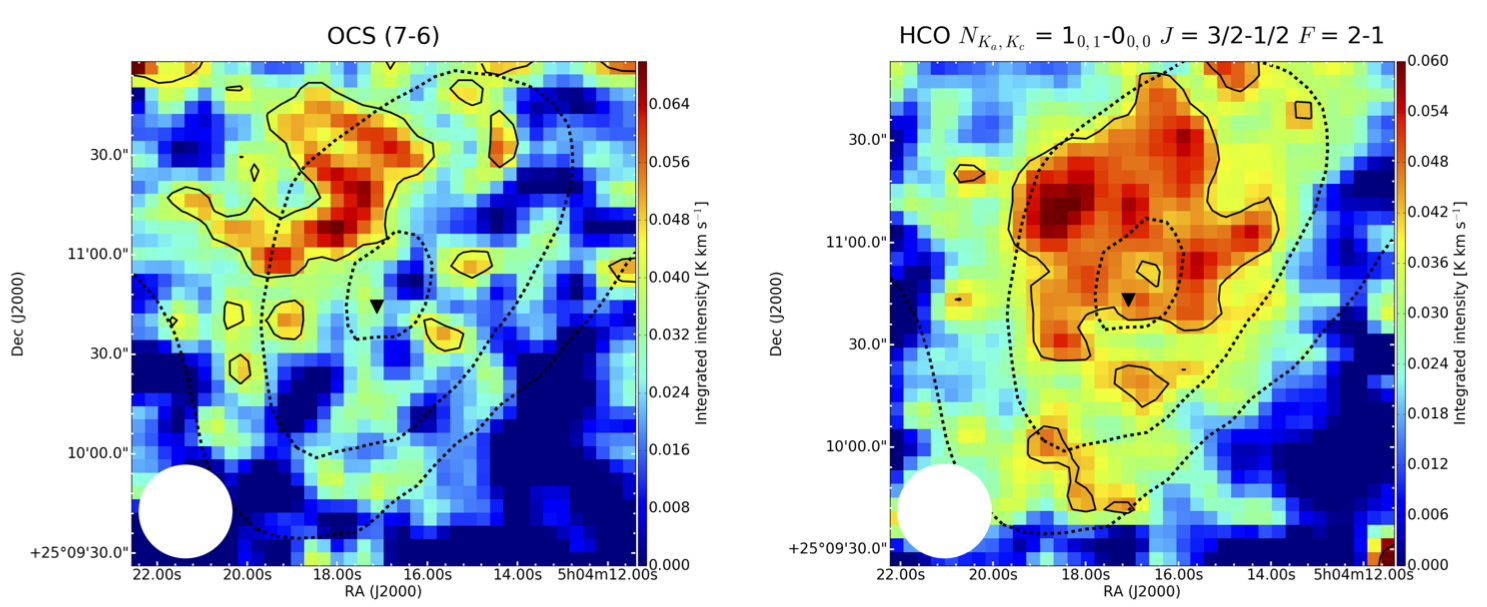}
  \caption{}
  \label{fig:}
\end{figure*}

\begin{figure*}[h]
 \centering
 \includegraphics [width=1\textwidth]{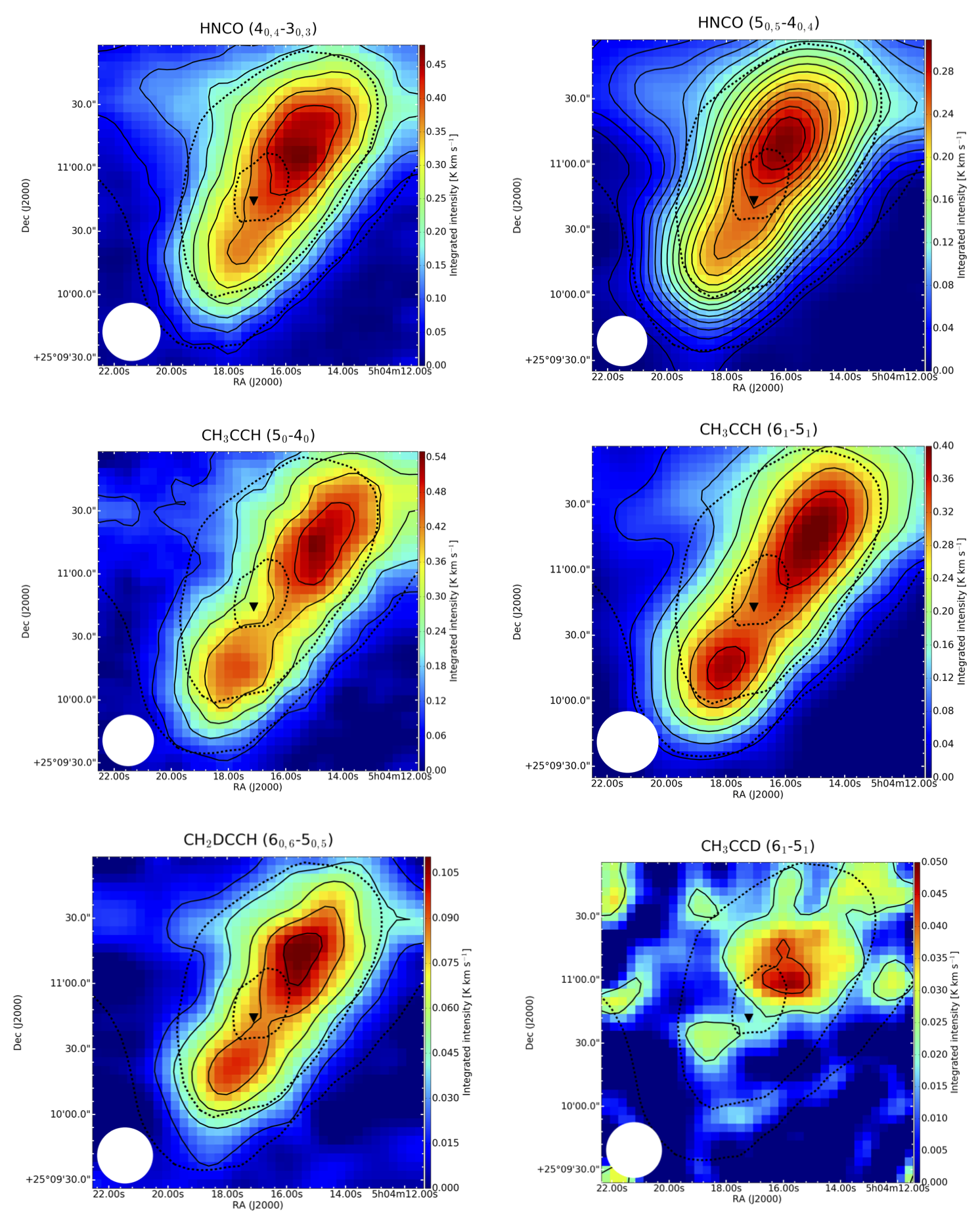}
  \caption{Maps of the molecules belonging to the HNCO peak family.}
  \label{fig:maps3}
\end{figure*}

\begin{figure*}[h]
 \centering
 \includegraphics [width=1\textwidth]{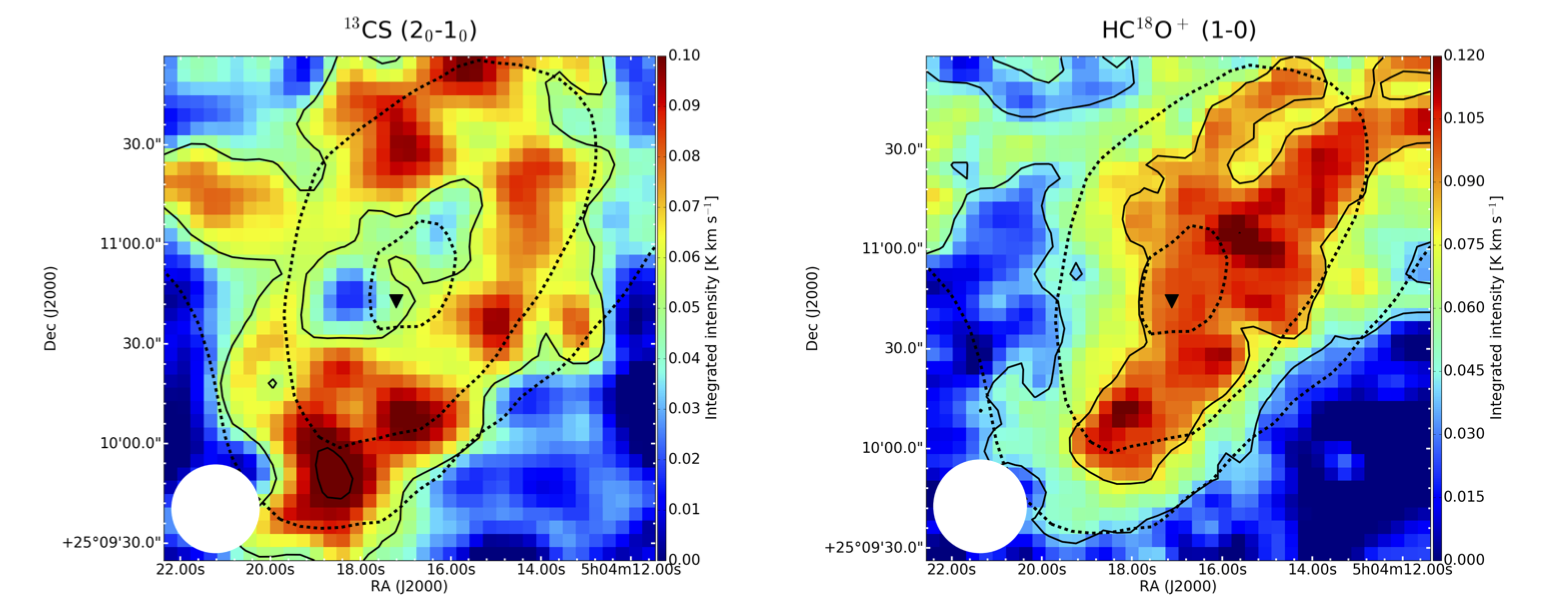}
  \caption{Maps of the molecules belonging to the category "Others".}
  \label{fig:maps4}
\end{figure*}

\clearpage

\begin{enumerate}
\item\textbf{The $c$-C$_3$H$_2$ family}\\
\begin{itemize}
\item{$c$-C$_3$H$_2$}
\end{itemize}
$c$-C$_3$H$_2$, cyclopropenylidene, is a three-membered carbon ring. Due to the presence of two unpaired electrons, it has a large dipole moment, 3.27 D \citep{lov92}, and consequently a very bright spectrum in the radio- and mm-band. Cyclopropenylidene was observed in the interstellar medium (ISM) before its detection was confirmed by laboratory work \citep{tha81, tha85}. Since then, it has been detected in a wide variety of sources \citep{mad89, nak11}.
$c$-C$_3$H$_2$ is an asymmetric top, therefore the labelling of its energy levels is $J_{K_a,K_c}$, with $J$ being the rotational quantum number and $K_a$ and $K_c$ the angular quantum numbers of the prolate and oblate symmetric top limits, respectively. We have mapped two lines of the normal species, $c$-C$_3$H$_2$ ($J_{K_a,K_c}$ = $3_{2,2} - 3_{1,3}$ and $J_{K_a,K_c}$ = $2_{0,2} - 1_{1,1}$ ) , and one of the $^{13}$C-species with the $^{13}$C off of the principal axis of the molecule ($J_{K_a,K_c}$ = 2$_{1,2}$ - 1$_{0,1}$).\\

\begin{itemize}
\item{H$_2$CCC ($l$-C$_3$H$_2$)}
\end{itemize}
H$_2$CCC, propadienylidene, is the linear and less stable isomer of cyclopropenylidene. Propadienylidene is usually not as widespread as its cyclic isomer. After its first detection in the laboratory \citep{vrt90}, it was detected in TMC-1, in IRC+10216 \citep{cer91, kaw91}, and in few other Galactic sources \citep{lis12}.  The cyclic-to-linear (hereafter $c/l$) ratio of C$_3$H$_2$ tends to increase with increasing $A_V$. Significant variations of this ratio are found in dense cores, ranging from 25 in TMC-1(CP) to 67 in TMC-1C \citep{spe16b, sip16}. We have mapped two rotational transitions of H$_2$CCC, namely the $J_{K_a,K_c}$ = 5$_{0,4}$ - 4$_{0,4}$ at 103.9 GHz, and the $J_{K_a,K_c}$ = 4$_{1,3}$ - 3$_{1,2}$ at 83.9 GHz.\\

\begin{itemize}
\item{C$_3$H}
\end{itemize}
The linear C$_3$H is a radical with a $^2$$\Pi$ electronic ground state. Given the presence of both the electronic orbital and the spin angular momentum, two ladders of rotational levels are present ($\Omega$ = 1/2 and 3/2), and they are described by Hund's "a" case. Here we have mapped two hyperfine transitions ($\Delta F$ = 1) in the $J$ = 9/2 - 7/2 transition of the $^2$$\Pi_{1/2}$ ground state. C$_3$H has been observed in IRC+10216 and TMC-1 \citep{tha85b} before its first laboratory detection \citep{got85, got86}, and it has subsequently been observed toward several sources, among those are translucent molecular clouds \citep{tur00}, a protoplanetary nebula \citep{par07}, and Sgr B2 \citep{mcg13}.\\

\begin{itemize}
\item{C$_4$H}
\end{itemize}
C$_4$H is a radical with $^2\Sigma^+$ electronic ground state, and its rotational spectrum is split in fine and hyperfine structure given the presence of electron and proton spin. We have mapped the $F$= 12 - 11 and 13 - 12 hyperfine lines (not resolved) of the $N$= 12-11 transition. C$_4$H was observed in the late 70s towards IRC+10216 \citep{gue78}before a laboratory spectrum was available, and its astronomical detection was later confirmed by the observations of lower rotational transitions in TMC-1 \citep{gue82}. Its first laboratory spectrum was detected in a Zeeman-modulated spectrometer by Gottlieb and coworkers \citep{got83}. It has now been detected in a wide range of sources, from photodissociation regions \citep{cua15} to dark clouds \citep{kai04}.\\

\begin{itemize}
\item{H$_2$CCO}
\end{itemize}
The microwave spectrum of ketene, H$_2$CCO, was observed in the 50s for the first time in the laboratory \citep{joh52}. In space, it has been detected in both dark and translucent clouds \citep{ohi91, tur99, tur77}. We have mapped the $J_{K_a,K_c}$ = 5$_{1,5}$ - 4$_{1,4}$ and $J_{K_a,K_c}$ = 5$_{1,4}$ - 4$_{1,3}$ transitions of H$_2$CCO.\\

\begin{itemize}
\item{HCCNC}
\end{itemize}
HCCNC, ethynyl isocyanide, is an isomer of HC$_3$N, cyanoacetylene. It was first observed in the laboratory in 1992 by means of Fourier transform microwave spectroscopy \citep{gua92} and observed in space by Kawaguchi et al. (1992) towards TMC-1. We have mapped one transition of HCCNC,  the $J$ = 9 - 8.\\

\begin{itemize}
\item{H$_2$CS}
\end{itemize}
The millimetre- and submillimetre-wave spectrum of thioformaldehyde was studied in the early 70s \citep{joh71}. It was observed in hot cores \citep{sin73}, dark clouds \citep{irv89}, and circumstellar envelopes \citep{agu08}. We present here the map of the  $J_{K_a,K_c}$ = 3$_{0,3}$ - 2$_{0,2}$ transition.\\
 
\begin{itemize} 
\item{HCS$^+$}
\end{itemize}
HCS$^+$ was observed in the ISM \citep{tha81} prior to its laboratory detection \citep{gud81}. It is a small cation with $^1$$\Sigma$$^+$ electronic ground state. We have mapped the $J$=2-1 rotational transition.\\

\begin{itemize}
\item{C$^{34}$S}
\end{itemize}
CS is a very abundant and widely distributed interstellar molecule. Towards starless cores, it presents the same abundance drop towards the centre as CO \citep{taf04}. We present here the map of the $J$ = 2 - 1 transition of $^{34}$SO towards L1544.\\

\begin{itemize}
\item{CCS}
\end{itemize}
Thioethenylidene, CCS, is a radical with electronic ground state $^3\Sigma^-$. Its rotational energy levels are described by the quantum numbers $J$ and $N$, referring respectively to the total angular momentum and to the angular momentum of the molecular frame, following Hund's "b" case. CCS was observed for the first time in the laboratory and in space in 1987 \citep{sai87}, and since then it has been observed in different kind of media, from dense to diffuse clouds. Furthermore, it was one of the first molecules mapped in L1544 \citep{oha99}. We have mapped the $N, J$ = 8, 7 - 7, 6 ; 7, 6 - 6, 5; 7, 7 - 6, 6; and 8, 9 - 7, 8 transitions. \\

\begin{itemize}
\item{CH$_3$CN}
\end{itemize}
Methyl cyanide was among the first molecules detected in the ISM \citep{sol71}. Given its high abundance, also its rarer isotopologues with $^{13}$C, D and $^{15}$N have been observed \citep{cum83, gue82, num98}. CH$_3$CN is prolate symmetric top rotor, hence its rotational levels are labeled $J_K$. We have mapped the $J_K$ = 6$_0$ - 5$_0$ rotational transition at 110 GHz.\\

\begin{itemize}
\item{HCC$^{13}$CN}
\end{itemize}
Cyanoacetylene is a linear molecule belonging to the family of cyanopolyyne (HC$_n$N with $n$ = 1, 3, 5, ..). Given the presence of low lying vibrational levels (bending modes), the vibrationally excited states of the normal isotopologue have also been detected in the ISM \citep{wyr03}. Here we present the map of the $J$ = 10 - 9 rotational transition.\\

\item\textbf{The dust peak family}\\
\begin{itemize}
\item{H$^{13}$CN}
\end{itemize}
HCN has been observed in a great variety of astrophysical environments, and several vibrationally excited states have been observed towards IRC+10216 \citep{cer11}. Being a linear molecule, it has a relatively simple spectroscopy and its rotational levels are described by the quantum numbers $J$ and $F$, because of the hyperfine structure due to the presence of the nitrogen atom. We mapped the $F$ =  2 - 1 hyperfine component of the $J$ = 1 - 0 rotational transition.\\

\begin{itemize}
\item{$^{13}$CN}
\end{itemize}
Cyanogen is a radical with $^2$$\Sigma$$^+$ electronic ground state. Like some other astronomically abundant ions and radicals, also CN has been observed in the ISM \citep{pen74} prior to its laboratory detection \citep{dix77}.
The spectrum of the $^{13}$CN isotopologue is characterised by a fine structure given by the interaction of the electron spin and the nuclear rotation, and a hyperfine structure rising from the presence of the $^{13}$C and the nitrogen atom. We have mapped the $N$ = 1 - 0, $F_1$ = 2 - 1, $F_2$ = 2 - 1, $F$ = 3 - 2 transition  of $^{13}$CN in L1544.\\

\begin{itemize}
\item{N$_2$H$^+$}
\end{itemize}
N$_2$H$^+$ is a molecular ion with $^1$$\Sigma$$^+$ electronic ground state. Its spectrum is complicated by the presence of two nitrogen atoms, and its $J$ = 1 - 0 transition is split in to seven hyperfine components \citep{cas95}. N$_2$H$^+$ has been widely observed in dense cores, and it has a crucial importance in astrochemistry because it is a proxy of N$_2$, not observable because it lacks dipole moment. Furthermore N$_2$H$^+$ doesn't suffer significant depletion like CO and, like other N-bearing molecules, is a very good tracer of the cold inner parts of starless cores. Here we mapped the $J$ = 1 - 0 $F_1$ = 0 - 1 $F$ = 1 - 2 hyperfine transition of N$_2$H$^+$. Maps of N$_2$H$^+$ and N$_2$D$^+$ have already been reported in Caselli et al. (2002).\\

\item\textbf{The methanol peak family}\\
\begin{itemize}
\item{CH$_3$OH}
\end{itemize}
Methanol is the smallest complex molecule in the ISM, and it has been extensively observed in dark clouds (e.g. \citealt{pra97}) and star forming regions (e.g. \citealt{bac96}). Given the presence of the CH$_3$ internal rotor, the resulting rotational-torsional spectrum is fairly complicated. Its torsional potential possesses three equivalent minima which lead to rotational transitions of symmetry A or E (doubly degenerated). Levels of different symmetry do not interact, hence the rotational transitions are labeled also with the symmetry state. In our case, we mapped the $J_{K_a,K_c}$ = 2$_{1,2}$ - 1$_{1,1}$ transition of the ($E_2$) state and the $J_{K_a,K_c}$ = 0$_{0,0}$ - 1$_{1,1}$ transition of the ($E_1$-$E_2$) state. This map has already been reported in \cite{biz14} and \cite{spe16}. \\

\begin{itemize}
\item{SO and $^{34}$SO}
\end{itemize}
SO is a radical with $^3\Sigma^-$ ground state. Its rotational energy levels are described by the quantum numbers $J$ and $N$, like for CCS. SO and $^{34}$SO have been extensively observed in star forming regions and also in extragalactic sources \citep{mar03}. Here we present the map of the $N, J$ = 2,2 - 1,1 and 3,2 - 2,1 lines of the main species, and the $N, J$ = 2,3 - 1,2 line for the $^{34}$SO isotopologue.\\

\begin{itemize}
\item{SO$_2$}
\end{itemize}
Sulfur dioxide is quite abundant in the ISM, in particular its lines together with the lines of methanol dominate the spectrum at submillimetre wavelenghts in star forming regions \citep{sut91}. SO$_2$ is a prolate asymmetric rotor, so its rotational energy levels are described by the quantum numbers $J_{K_a,K_c}$ . We have mapped the $J_{K_a,K_c}$ = 3$_{1,3}$ - 2$_{0,2}$ rotational transition.\\

\begin{itemize}
\item{OCS}
\end{itemize}
OCS is a linear molecule with a simple spectrum with transitions every $\sim$12 GHz. It was detected for the first time in the ISM in the early 70s \citep{jef71} and it has been so far observed in a wide variety of media. Furthermore, it is the only S-bearing molecule detected in interstellar ices \citep{bog15}. Here we present the map the of $J$=7-6 transition at 85.139 GHz.\\

\begin{itemize}
\item{HCO}
\end{itemize}
The formyl radical, HCO has been observed towards both diffuse and dense clouds \citep{lis14, agu15}. The interaction of the electron spin with the rotation of the molecule splits each rotational level into a doublet. An additional split is caused by the magnetic interaction of the electron spin with the hydrogen nuclear spin. The $N_{K_a,K_c}$ = 1$_{0,1}$ - 0$_{0,0}$ rotational transition has 6 hyperfine components, here we present the map of one hyperfine component, namely the $J$ = 3/2 - 1/2 $F$ = 2 - 1.\\

\item\textbf{The HNCO peak family}\\
\begin{itemize}
\item{CH$_3$CCH, CH$_2$DCCH and CH$_3$CCD}
\end{itemize}
Propyne (methyl acetylene) is a stable hydrocarbon, whose spectrum has long been studied in the laboratory \citep{mul00} and in space \citep{sny73, irv81}. While the normal species is a symmetric rotor, and its rotational levels can be labeled with the $J$ and $K$ quantum numbers, the substitution of one hydrogen with deuterium in the methyl group (CH$_3$), breaks the symmetry and three quantum numbers are necessary to describe the spectrum of CH$_2$DCCH. We have mapped the $J_K$ = 5$_0$-4$_0$ line of CH$_3$CCH and the  $J_K$ = 6$_1$ - 5$_1$ line of CH$_3$CCD , and the $J_{K_a,K_c}$ = 6$_{0,6}$ - 5$_{0,5} $ line of CH$_2$DCCH.\\

\begin{itemize}
\item{HNCO}
\end{itemize}
Isocyanic acid (HNCO) has been vastly studied both in the laboratory and in the ISM. Also its isomer, HCNO, has been observed in space \citep{mar09}, and very recently its deuterated isotopologue \citep{cou16}. Gas-phase models fail to reproduce the observed abundances of HNCO, and the formation on icy mantels of dust grains has been proposed and tested experimentally \citep{jim14}. HNCO has been linked to the formation of formamide (NH$_2$CHO), a prebiotic interstellar molecule \citep{lop15}. We have mapped the $J_{K_a,K_c}$ = 4$_{0,4}$ - 3$_{0,3}$ and $J_{K_a,K_c}$ = 5$_{0,5}$ - 4$_{0,4}$ lines of HNCO. \\

\item\textbf{Others}\\
\end{enumerate}
\begin{itemize}
\item{$^{13}$CS}
\end{itemize}
We have mapped the $J$ = 2 - 1 transition of $^{13}$CS. While the less abundant isotopologue C$^{34}S$ belongs to the $c$-C$_3$H$_2$ family, the $^{13}$C isotopologue is more abundant and shows a diffuse distribution, it hence does not belong to any of the above mentioned categories.\\

\begin{itemize}
\item{HC$^{18}$O$^+$}
\end{itemize}
HCO$^+$, protonated carbon monoxide, was observed in 1970 towards several high mass start forming regions by \citep{buh70}. The carrier of this line with "unknown extraterrestrial origin" was proposed to be HCO$^+$ by \cite{kle70}. The detection was not definitive until the first laboratory detection of HCO$^+$ \citep{woo75}.  The maps of H$^{13}$CO$^+$ ($J$ = 1 - 0) and DCO$^+$ ($J$ = 2 - 1 and 3 - 2) have already been presented in Caselli et al. (2002). Here we present the map of the $J$ = 1 - 0 of HC$^{18}$O$^+$.

\section{Principal component analysis}
\begin{figure*}
 \includegraphics [width=0.9\textwidth]{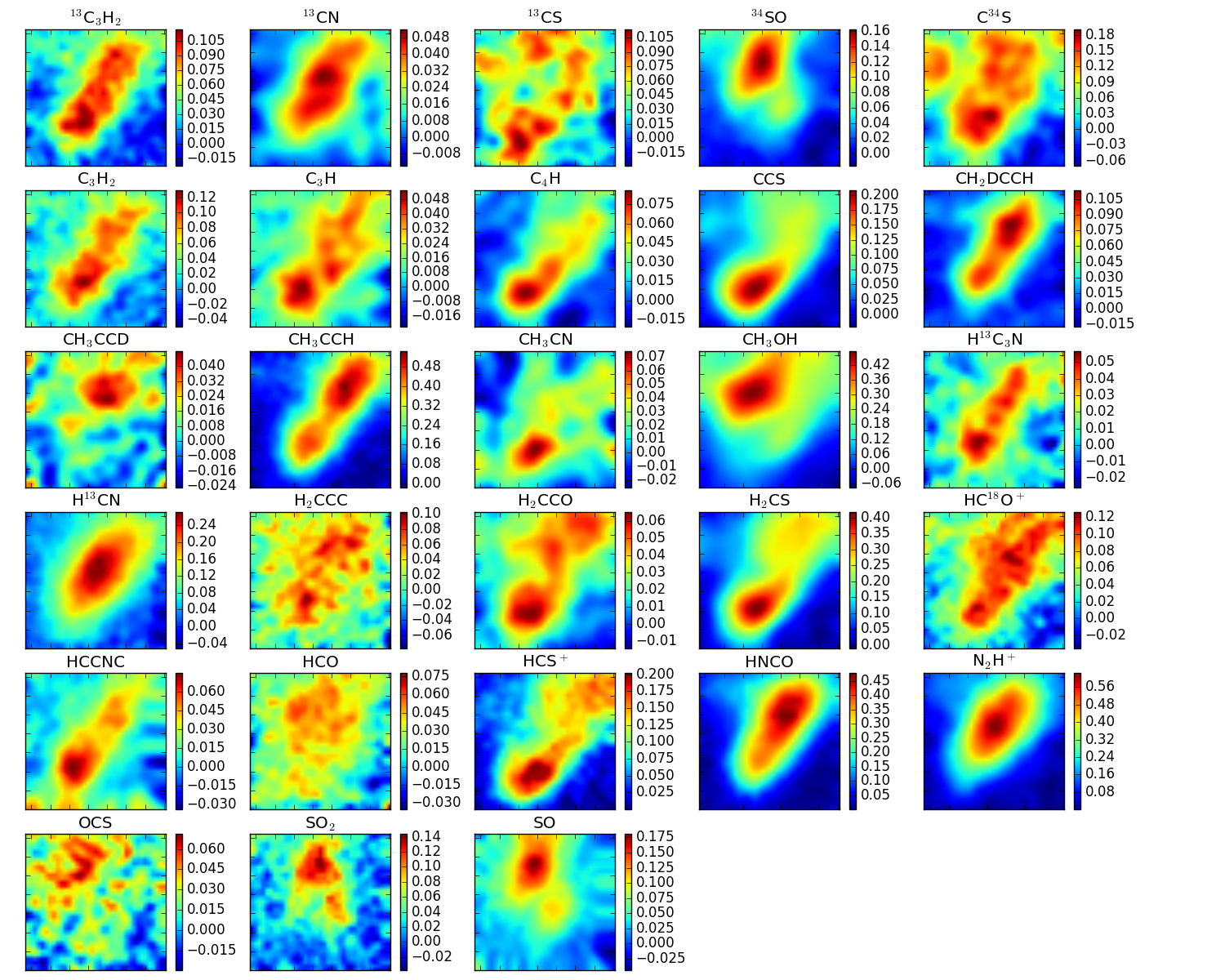}
 \caption{Raw dataset. The colour scale shows the integrated emission in K km s$^{-1}$}
  \label{fig:10}
\end{figure*}

\begin{figure*}
 \includegraphics [width=0.9\textwidth]{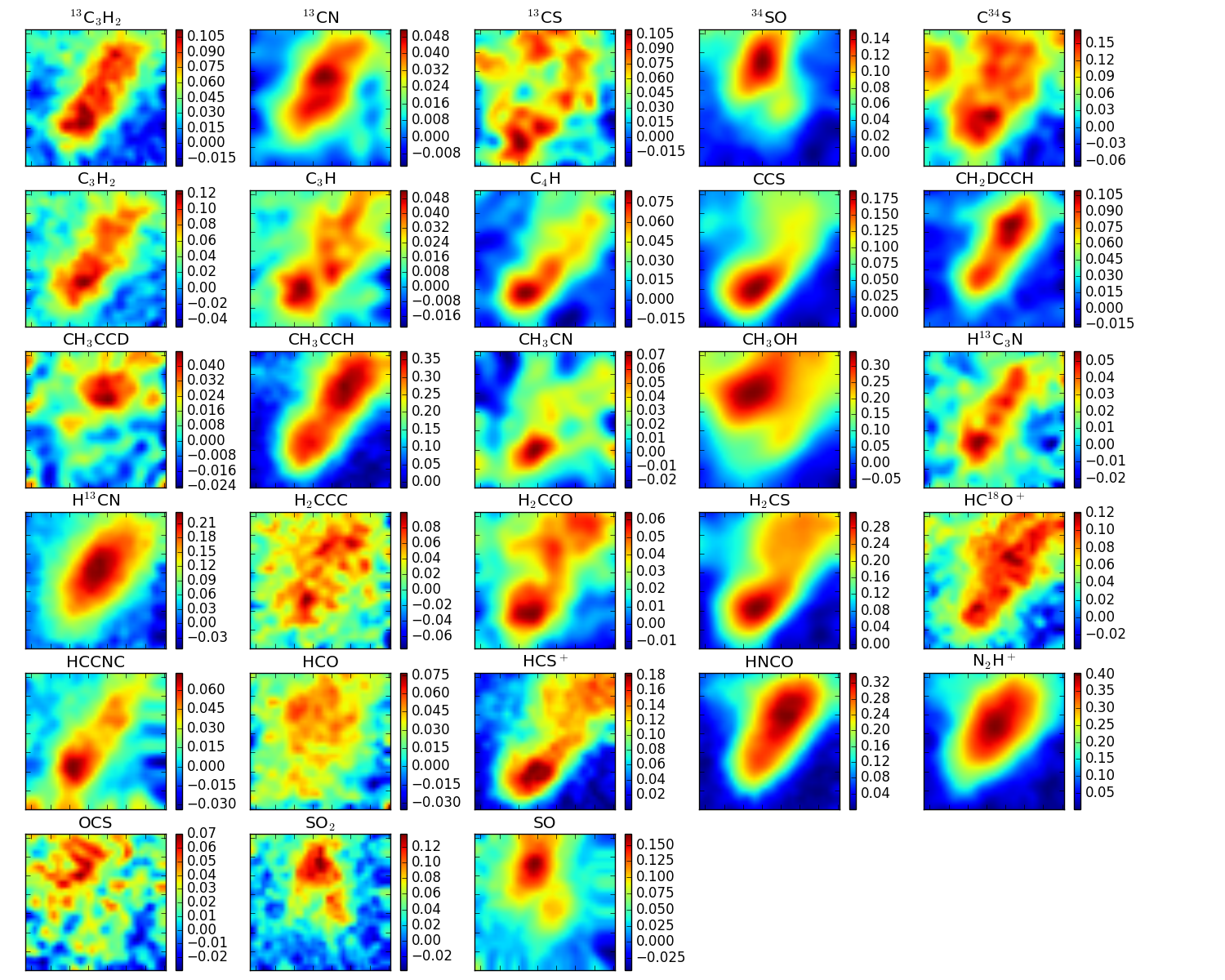}
 \caption{Dataset after the reparametrisation following the method reported in \cite{gra17}. The colour scale shows the integrated emission in K km s$^{-1}$}
  \label{fig:11}
\end{figure*}

\begin{figure*}
 \includegraphics [width=0.9\textwidth]{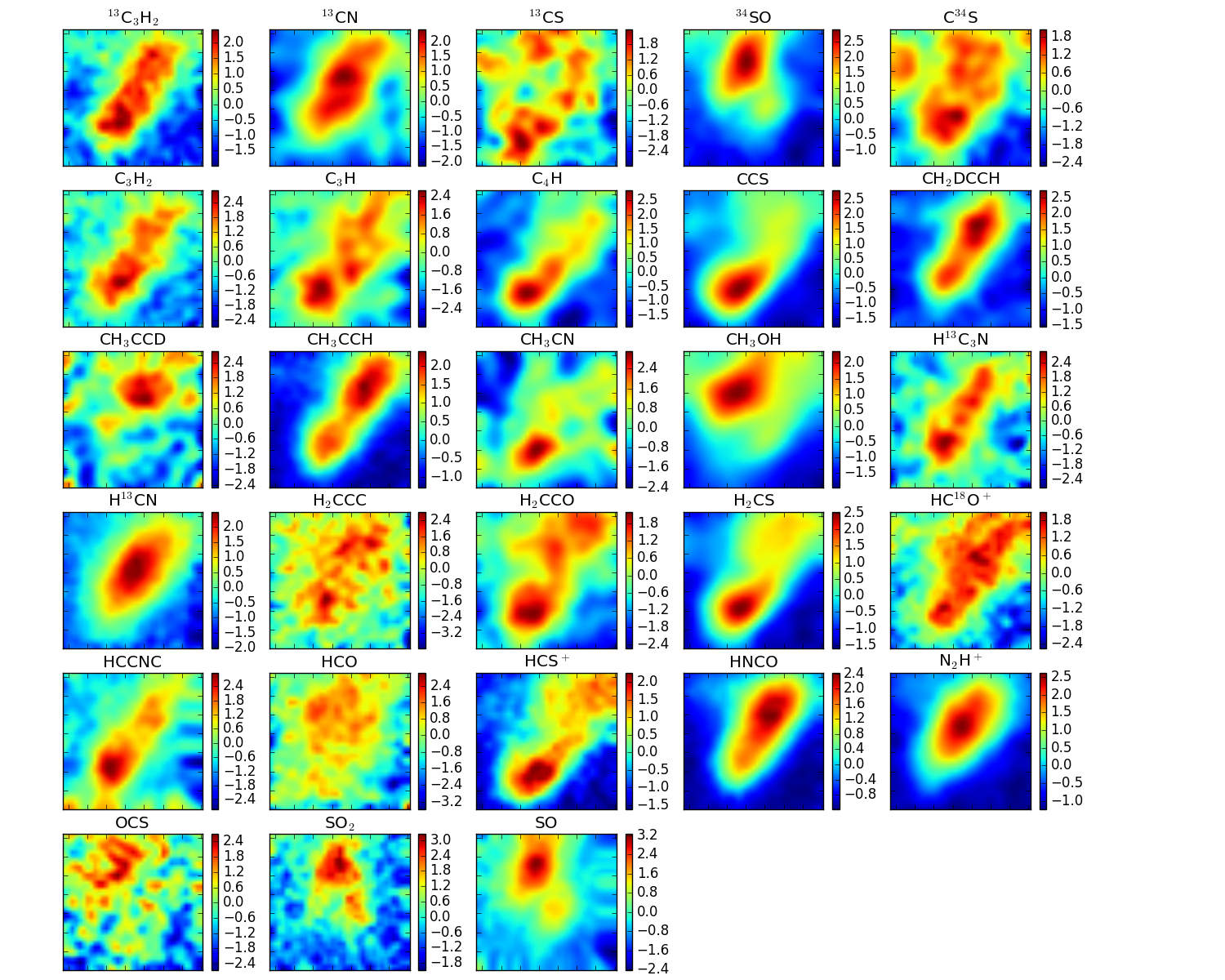}
 \caption{Dataset after the standardisation following the method reported in \cite{ung97}. The colour scale shows the integrated emission in K km s$^{-1}$}
  \label{fig:12}
\end{figure*}

\begin{figure*}
 \centering
 \includegraphics [width=1\textwidth]{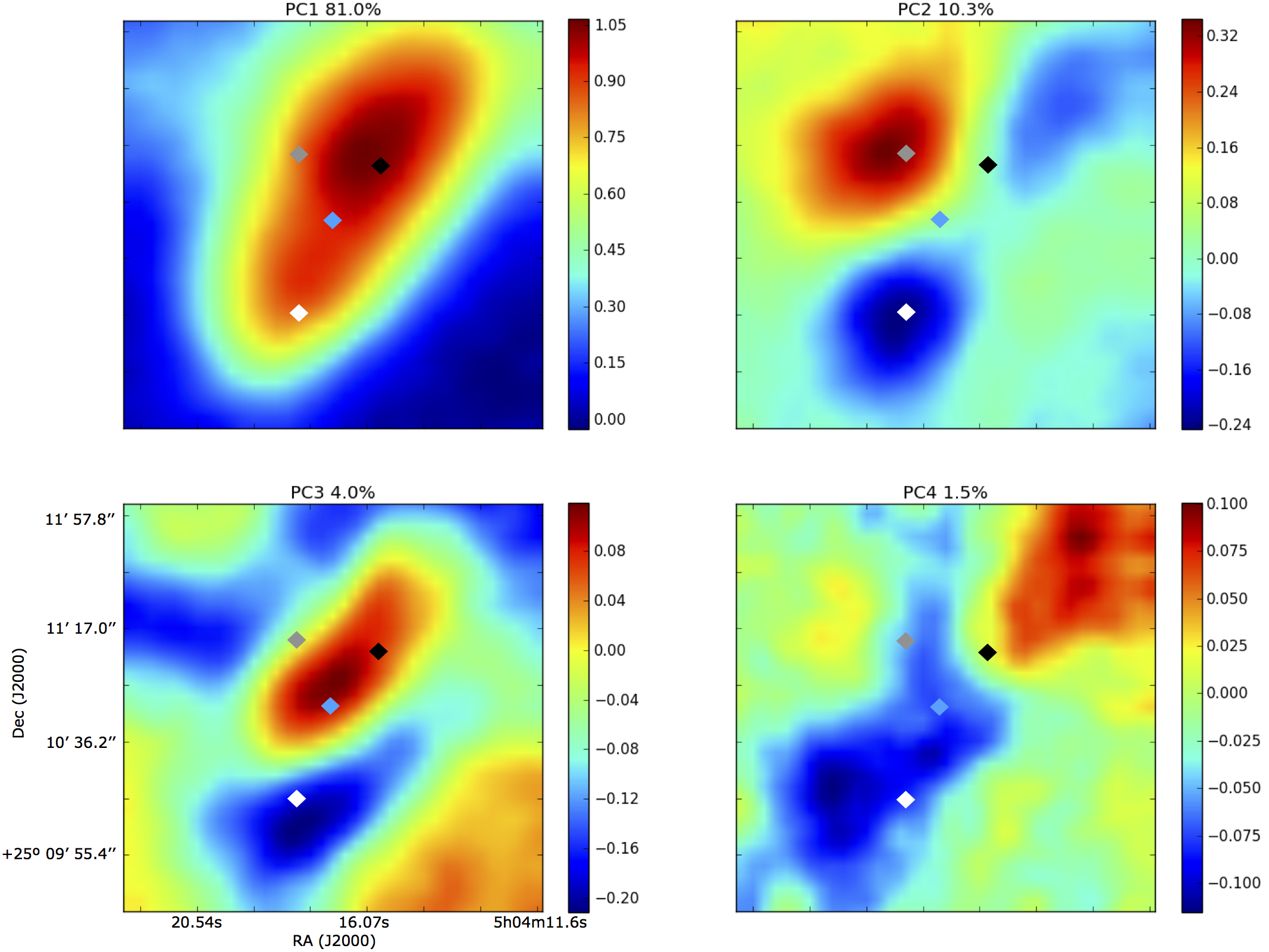}
 \caption{Maps of the first four principal components obtained by performing the PCA on the raw data. The blue, black, white and grey diamonds indicate the dust, the HNCO, the $c$-C$_3$H$_2$, and the methanol peaks respectively. }
  \label{fig:1}
\end{figure*}

\begin{figure}
 \centering
 \includegraphics [width=0.5\textwidth]{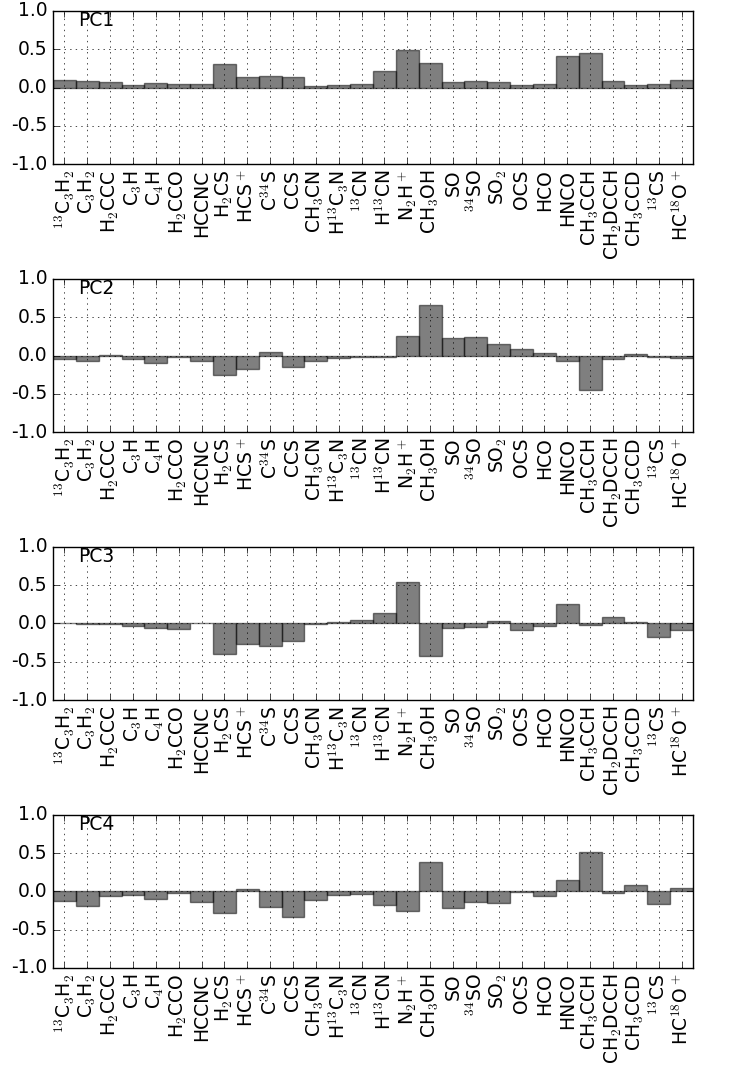}
 \caption{Contribution of each molecule to the first four PC, obtained by performing the PCA on the raw data.}
  \label{fig:2}
\end{figure}

\begin{figure}
 \centering
 \includegraphics [width=0.5\textwidth]{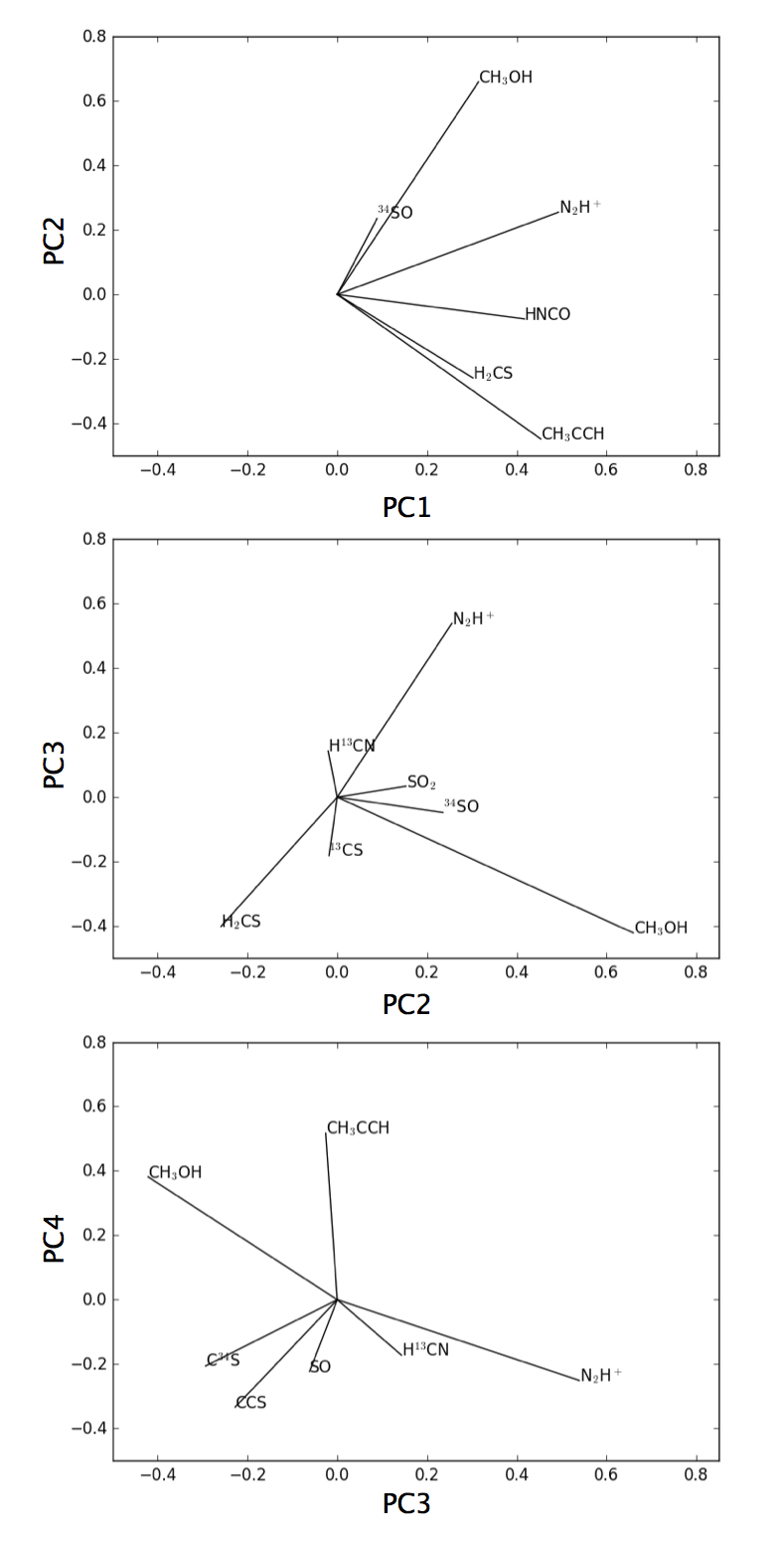}
 \caption{Correlation wheels where each molecule has as coordinates their correlation coefficients to each PC, obtained by performing the PCA on the raw data.}
  \label{fig:3}
\end{figure}

\begin{figure*}
 \centering
 \includegraphics [width=1\textwidth]{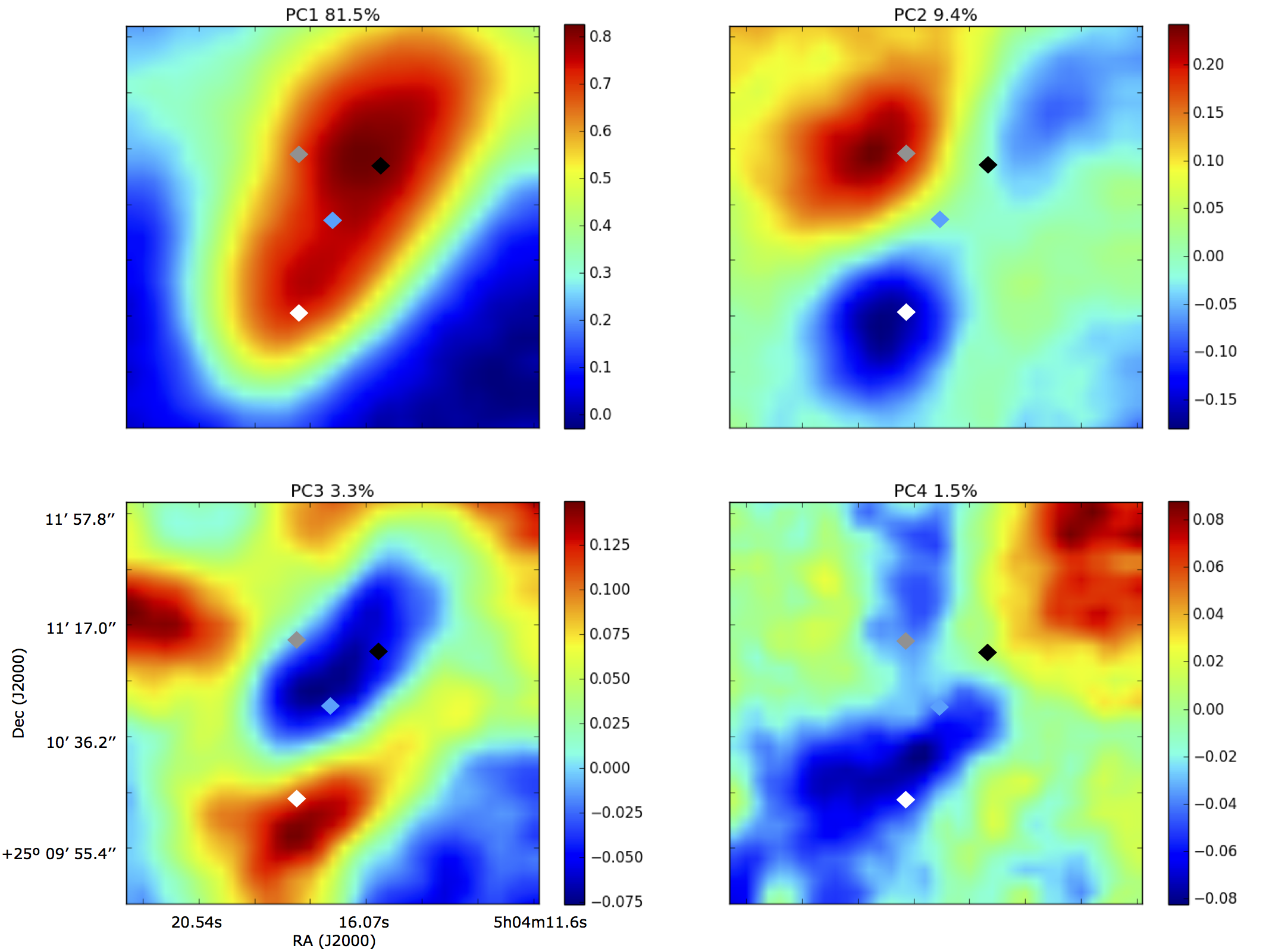}
 \caption{Maps of the first four principal components obtained by performing the PCA on data reparametrised as in Gratier et al. 2017. The change of sign of PC3 and PC5 with respect to the maps in Figure \ref{fig:1} depends on the value of the parameter $a$, which in our case is arbitrarily chosen (see text). The blue, black, white and grey diamonds indicate the dust, the HNCO, the $c$-C$_3$H$_2$, and the methanol peaks respectively.}
  \label{fig:4}
\end{figure*}

\begin{figure}
 \centering
 \includegraphics [width=0.5\textwidth]{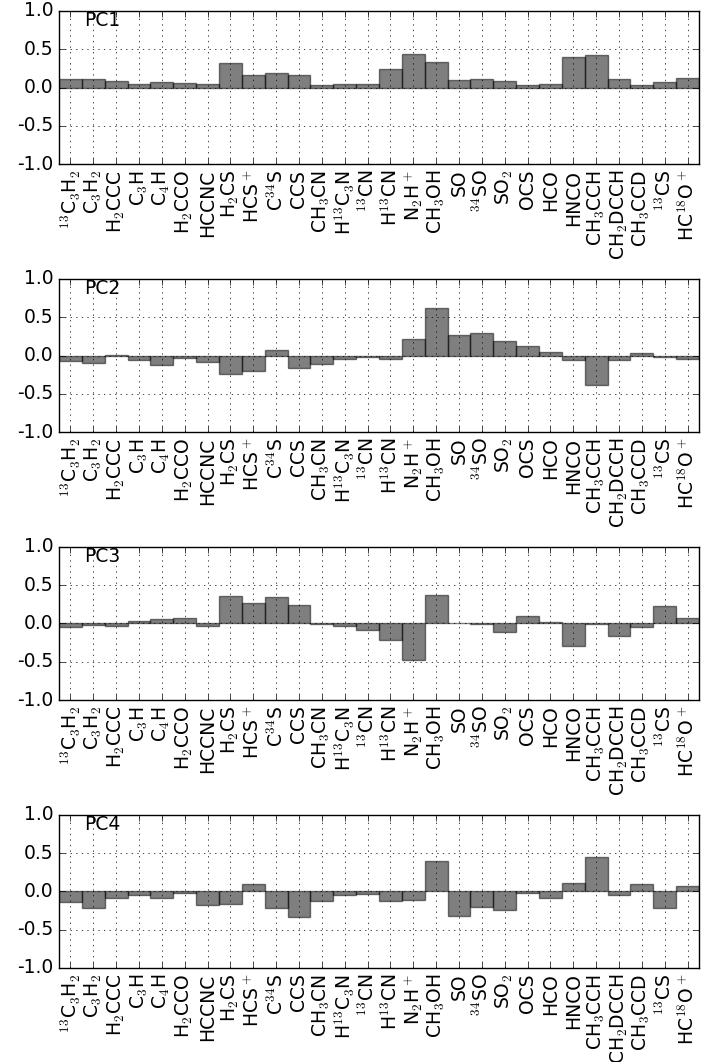}
 \caption{Contribution of each molecule to the first four PC, obtained by performing the PCA on data reparametrised as in Gratier et al. 2017.}
  \label{fig:5}
\end{figure}

\begin{figure}
 \centering
 \includegraphics [width=0.5\textwidth]{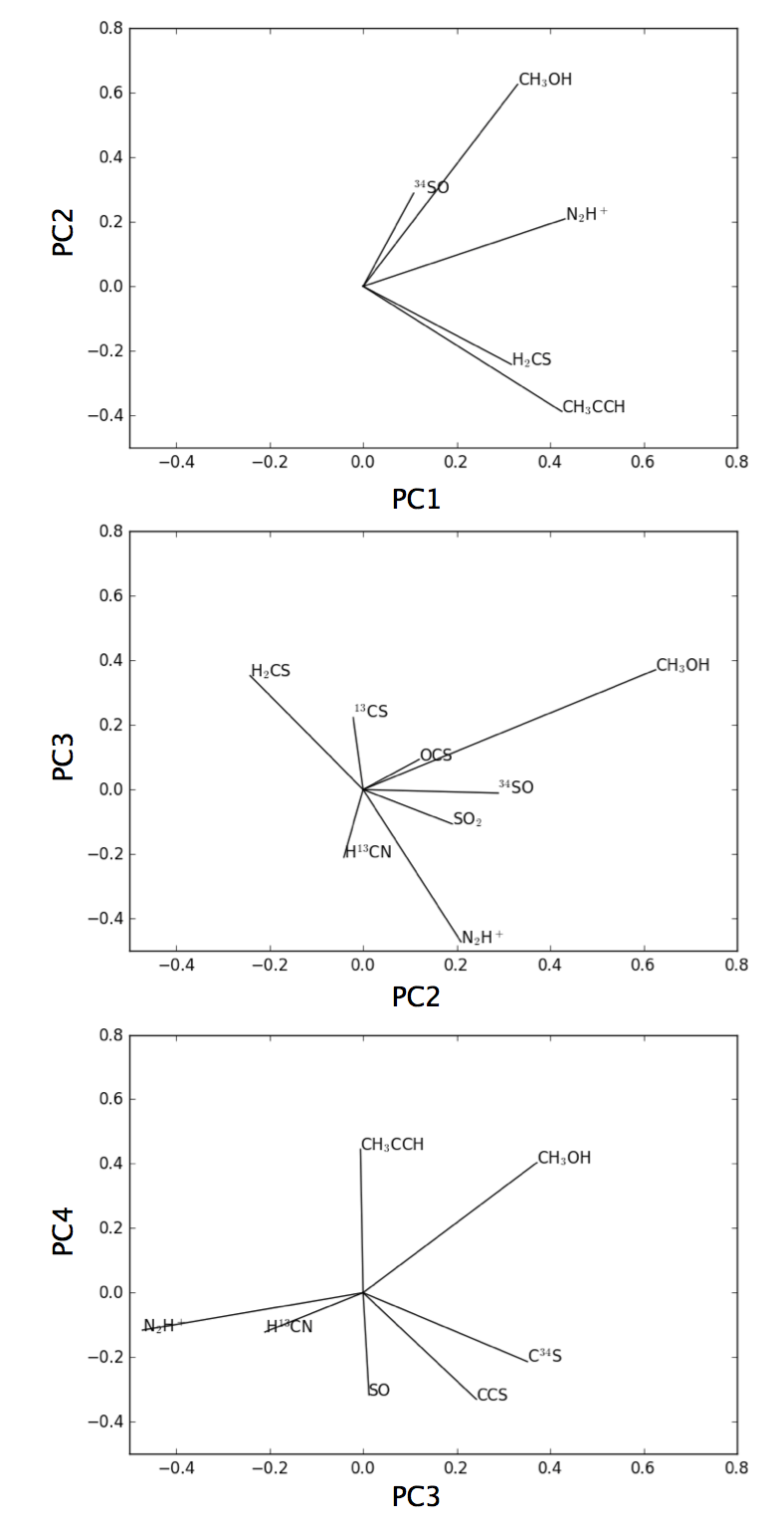}
 \caption{Correlation wheels where each molecule has as coordinates their correlation coefficients to each PC, obtained by performing the PCA on data reparametrised as in Gratier et al. 2017.}
  \label{fig:6}
\end{figure}

In order to decide which was the most rigorous way to carry the principal component analysis on our dataset, we have performed it on the raw data, on the standardised data as in \cite{ung97}, and on the reparametrised data as in \cite{gra17}, and compared the different results.\\
In \cite{gra17} the reparametrisation has been done using the following formula:

\begin{eqnarray*}
T(x) &=& a \times\mathrm{asinh} (x/a),\\
\end{eqnarray*}

\noindent where the $a$ parameter is 8$\times$0.08 K, with 0.08 K being the median of the noise among all the maps in his dataset. The parameter $a$ was optimised so that it would maximise the correlation (quantified with the Spearman's correlation coefficient) of the PC maps with three physical maps of the region, namely the H$_2$ column density, the volume density and the UV illumination. In our case we could not optimise the $a$ parameter as done in \cite{gra17} because the value of the Spearman's correlation coefficient of the physical maps of L1544 with the PC maps does not change substantially when varying the parameter $a$. The results provided here on the reparametrised data use $a$ = 15 $\times$ 0.016 K, where 0.016 K is the median of the noise of our maps.\\
The PCA gives an unbiased result when performed on data that have a gaussian distribution. 
The maps before and after the reparametrisation and standardisation are shown in Figures \ref{fig:10} - \ref{fig:12}.
The results of the analysis on the raw data and on the reparametrised data are reported in Figures \ref{fig:1} - \ref{fig:6}.
When comparing the PC maps obtained on the raw data and on the data that have been reparametrised or standardised, we can see that there is not a substantial difference in their shapes. In contrast with what has been done in previous works, we do not include in our dataset the maps of very bright lines ($^{12}$CO for example), and hence the dynamic range in our dataset is not as large. As a consequence, calculating the eigenvectors and eigenvalues on the covariance or correlation matrix does not produce results which are very different from each other.
Still, there are some differences and they need to be taken into account in order to do a rigorous analysis of the results.
Both in the case of the raw data and with the reparametrisation from \cite{gra17}, the first PC map accounts for $\sim$ 80$\%$ of the total correlation and peaks towards the HNCO peak, while in the standardised data the first PC map accounts for $\sim$ 65$\%$ and it peaks toward the $c$-C$_3$H$_2$ and HNCO peak. This difference can be explained by the fact that the brightest lines in our sample are those from HNCO and CH$_3$CCH, they peak towards the HNCO peak and they dominate the results in the case of performing the PCA on the raw data and on the reparametrised data. This is clear if we look at the emission maps before and after the reparametrisation and the standardisation in Figures \ref{fig:10} - \ref{fig:12}. While the reparametrised data already show a decrease in the intensity of the emission of the brightest lines (i.e. CH$_3$OH and CH$_3$CCH), it is just when the data are standardised that the differences between bright and weak lines are equalised. For these reasons we have decided to focus our discussion on the results of the analysis performed on the standardised data.


\clearpage

\end{appendix}

\end{document}